\documentclass{article}
\usepackage{amsmath}
\usepackage{amssymb}
\usepackage{amsfonts}
\usepackage{bm}
\usepackage[dvips,xdvi]{graphicx}
\usepackage{cite}
\begin{document}



\title{
{\bf Single Chain Slip-Spring Model
for Fast Rheology Simulations
of Entangled Polymers on GPU}}
\author{Takashi Uneyama\footnote{
E-mail: uneyama@scl.kyoto-u.ac.jp,
Tel: +81-774-38-3147, Fax: +81-774-38-3139} \\
\\
JST-CREST, Institute for Chemical Research, Kyoto University, \\
Gokasho, Uji 611-0011, Japan}

\date{}

\maketitle


\begin{abstract}


We propose a single chain slip-spring model, which is based on the
slip-spring model by Likhtman [A. E. Likhtman, {\it Macromolecules}, {\bf 38},
6128 (2005)], for fast rheology simulations of entangled polymers on a
GPU. We modify the original slip-spring model slightly
for efficient calculations on a GPU. Our model is designed to satisfy
the detailed balance condition, which enables us to analyze its static
or linear response properties easily. We theoretically analyze several
statistical properties of the model, such as the linear response, which
will be useful to analyze simulation data.
We show that our model can reproduce several
rheological properties such as the linear viscoelasticity or the
viscosity growth qualitatively. We also show that the use of a GPU
can improve the performance drastically.
\end{abstract}

\vspace{\baselineskip}



\section{Introduction}

Polymeric liquids exhibit various interesting macroscopic flows.
To simulate macroscopic flow behaviors of entangled polymeric materials,
we need to incorporate microscopic or mesoscopic polymer models (which
reproduces required rheological properties) with the
macroscopic fluid model (such as the Cauchy equation).
Considering the computational costs, phenomenological constitutive
equation models
\cite{Doi-Edwards-book,Ianniruberto-Marrucci-2001,Marrucci-Ianniruberto-2002,Likhtman-Graham-2003}
are reasonable for macroscopic simulations. However, most of constitutive equation
models involve rather rough or physically unclear approximations, which
is not fully justified.
To avoid such uncertainties, we can use microscopic or
mesoscopic molecular models
\cite{Kremer-Grest-1990,Hua-Schieber-1998,Masubuchi-Takimoto-Koyama-Ianniruberto-Greco-Marrucci-2001,Doi-Takimoto-2003,Schieber-Neergaard-Gupta-2003,Likhtman-2005,Nair-Schieber-2006,Kindt-Briels-2007}.
Although these molecular models also involve approximations, generally
they require less phenomenological parameters, and we expect they are
more precise than constitutive equation models.
Thus we expect the use of molecular
models instead of constitutive equation models can improve the
macroscopic flow simulation.

Several different methods have been proposed to incorporate the
mesoscopic polymer models with macroscopic fluid models.
For example, the CONNFFESSIT type methods \cite{Laso-Ottinger-1993},
the Lagrangian particle based methods \cite{Halin-Lielens-Keunings-Legat-1998,Murashima-Taniguchi-2010},
or recently developed finite volume based hybrid method \cite{Yasuda-Yamamoto-2009,Yasuda-Yamamoto-2010}
were proposed and achieved success to simulate macroscopic flows of
polymers.
{ Because these models directly combine different models which
have different time and length scales (microscopic or mesoscopic
rheological model, and macroscopic fluid model), we call them
``multiscale'' simulation models in this work.}
In these multiscale models, microscopic or mesoscopic rheological
simulations are
directly used to calculate macroscopic quantity such as the stress tensor,
instead of phenomenological constitutive
equation models.
In other words, microscopic or mesoscopic simulations
are embedded to macroscopic fluid elements.
Clearly such simulations require many
microscopic or mesoscopic rheological simulations which are numerically not
efficient. { To simulate large scale and/or long time
macroscopic phenomena, therefore
it is demanding to perform many microscopic or mesoscopic simulations
efficiently.}

One possible way is to use hardwares for acceleration. So far, several
acceleration hardwares (such as MDGRAPE \cite{mdgrape-url,Susukita-Ebisuzaki-Elmegreen-Furusawa-Kato-Kawai-Kobayashi-Koishi-McNiven-Narumi-Yasuoka-2003,Narumi-Ohno-Futatsugi-Okimoto-Suenaga-Yanai-Taiji-2006}, ClearSpeed \cite{clearspeed-url,clearspeed-software-development-kit-reference-manual},
or Cell Broadband Engine \cite{cell-scei-url,cell-ibm-url,cellbe-programming-handbook})
have been developed and utilized to accelerate calculations.
Recently, the acceleration by a graphic processor unit (GPU) \cite{cuda-url,Lindholm-Nickolls-Oberman-Montrym-2008,nvidia-cuda-programming-guide,atistream-url,ati-stream-computing-user-guide},
which is called general purpose GPU (GPGPU) programming \cite{Owens-Luebke-Govindaraju-Harris-Kruger-Lefohn-Purcell-2007}, is utilized
to accelerate various calculations including molecular dynamics \cite{Hamada-Iitaka-2007,Anderson-Lorenz-Travesset-2008}, fluid
dynamics \cite{Cohen-Garland-2009}, Monte Carlo models
\cite{Preis-Virnau-Paul-Schneider-2009}, or stochastic differential
equation models \cite{Januszewski-Kostur-2010}.
It is reported that GPUs can accelerate
simulations drastically, although the efficiency strongly depends on a
target. Because GPUs are much cheaper than other
acceleration hardwares and new GPUs are developed continuously, they are
considered to be promising compared with other previous acceleration hardwares.
Thus we expect that rheological simulations
on a GPU will be a good candidate for the embedded fast rheological simulation.

In this work we employ the slip-spring model, which is originally
proposed by Likhtman \cite{Likhtman-2005} to study the structure
and dynamics of entangled polymers, for simulations on
a GPU. We propose a slightly modified, single chain version of the
slip-spring model. We design our model to be suitable for simulations
on a GPU. We also design our model to fully satisfy the detailed balance
condition. We derive several statistical properties such as equilibrium
probability distributions or a formula for the relaxation modulus.
By performing rheological simulations both on a CPU and on a GPU, we
study rheological properties of the model and the acceleration effect by
a GPU. It is shown that our model reproduces rheological
properties reasonably and it enables efficient calculations on a GPU.

\section{Model}

Although there are various mesoscopic molecular models to calculate rheological
properties of entangled polymers
\cite{Kremer-Grest-1990,Hua-Schieber-1998,Nair-Schieber-2006,Masubuchi-Takimoto-Koyama-Ianniruberto-Greco-Marrucci-2001,Doi-Takimoto-2003,Schieber-Neergaard-Gupta-2003,Likhtman-2005,Kindt-Briels-2007},
not all models are suitable for simulations on a GPU.
In this work, we employ the slip-spring model proposed by Likhtman
\cite{Likhtman-2005}.
In the Likhtman's original slip-spring model, polymer
chains are expressed as ideal non-interacting Rouse chains and the
entanglement effect is mimicked by slip-springs. One end of a
slip-spring is fixed in space and another end is attached to the polymer
chain. As we will show later, such a model is suitable for calculations
on a GPU.

In this section, we show a single chain slip-spring model, which is a
variant of the original slip-spring model.
We modify the Likhtman's model slightly to make it suitable for
simulations on a GPU. To implement simulation programs on a GPU, we need
to use GPU specific programming environment and it has several
limitations. Thus we design our dynamics model to be as simple as possible.
At the same time, to make the model physically natural and to make
static properties simple, we attempt to make the dynamics model to satisfy
the detailed balance condition.
We also discuss about the equilibrium
statistical properties and the linear response theory of the slip-spring
model, which is useful to calculate the linear viscoelasticity.

\subsection{Free Energy, Grand Potential, and Equilibrium Statistics}

We first describe the model of our single chain slip-spring model.
We model an entangled polymer chain by an ideal Rouse chain and slip-springs.
The conformation of a polymer chain is expressed by positions of beads
$\lbrace \bm{R}_{i} \rbrace$, where $i = 1,2,\dots,N$ is the bead index
($N$ is the total number of beads). The conformation of slip-springs are
expressed by two set of variables; the bead indices of slipl-spring $\lbrace S_{j}
\rbrace$ and the anchoring point positions $\lbrace \bm{A}_{j}
\rbrace$, where $j = 1,2,\dots,Z$ with $Z \ge 0$ being the total number of
slip-springs.
One end of the $j$-th slip-spring is attached to the $S_{j}$-th bead of
a polymer chain, and another end is anchored in space, at $\bm{A}_{j}$.
We assume that $S_{j}$ is an integer value and $1 \le
S_{j} \le N$.
The total number of slip-springs $Z$ is not constant because
slip-springs can be spontaneously constructed or destructed. Thus variables
$\lbrace \bm{R}_{i} \rbrace$, $\lbrace \bm{A}_{j} \rbrace$,
$\lbrace S_{j} \rbrace$, and $Z$ are required to specify a state uniquely.

To make the statistical mechanical properties of the model clearly,
we first describe the free
energy of a single chain with slip-springs.
The free energy of a polymer chain and attached slip-springs can be expressed as follows.
\begin{equation}
 \label{free_energy_single_chain}
  \mathcal{F}(\lbrace \bm{R}_{i} \rbrace,\lbrace \bm{A}_{j} \rbrace,\lbrace
  S_{j} \rbrace, Z) = \sum_{i = 1}^{N - 1} \frac{3 k_{B} T}{2 b^{2}}
  (\bm{R}_{i + 1} -
  \bm{R}_{i})^{2}
  + \sum_{j = 1}^{Z} \frac{3 k_{B} T}{2 N_{s} b^{2}}
  (\bm{R}_{S_{j}} - \bm{A}_{j})^{2}
\end{equation}
where $k_{B}$ is the Boltzmann constant, $T$ is the temperature and $b$
is the bead size (segment size).
$N_{s}$ is a parameter which represents the strength of the
slip-springs. As mentioned, the total number of slip-springs, $Z$, is
not constant but fluctuates in time
since slip-springs are dynamically constructed or destructed.
To handle such a variable, it is convenient to use grand canonical type
ensemble for slip-springs, which is originally introduced by Schieber
\cite{Schieber-2003} for the slip-link model.
The grand potential of the system can be expressed as follows.
\begin{equation}
 \label{grand_potential_single_chain}
  \begin{split}
   \mathcal{J}(\lbrace \bm{R}_{i} \rbrace,\lbrace \bm{A}_{j} \rbrace,\lbrace
   S_{j} \rbrace, Z)
   & = \mathcal{F}(\lbrace \bm{R}_{i} \rbrace,\lbrace \bm{A}_{j} \rbrace,\lbrace
   S_{j} \rbrace, Z)
   - \nu Z
  \end{split}
\end{equation}
where $\nu$ is the effective chemical potential for a slip-spring. As
long as the detailed balance condition is satisfied, all the equilibrium
statistical properties can be calculated from the grand potential
\eqref{grand_potential_single_chain}.
As we discuss later, we can design dynamics to satisfy the detailed
balance condition, and thus all the equilibrium properties shown in the
followings can be also applied to our dynamics model.

Before performing numerical simulations, we analytically show several
equilibrium statistical
properties of our single chain slip-spring model. In this section, we
show equilibrium and linear-response properties.
Here we may emphasize that the results in this section is based on the
usual equilibrium statistical physics and the linear response theory in
non-equilibrium statistical physics. As long as the detailed balance
condition holds, all the results in this section also hold.

By using the grand potential \eqref{grand_potential_single_chain}, we can calculate the
grand partition function of the system.
\begin{equation}
 \label{grand_partition_function_single_chain}
  \begin{split}
  \Xi & \equiv \sum_{Z = 0}^{\infty} \frac{1}{\Lambda^{3 N} \Lambda_{s}^{3 Z} Z!} \sum_{\lbrace S_{j} \rbrace}
  \int d\lbrace \bm{R}_{i} \rbrace d\lbrace
  \bm{A}_{j} \rbrace \,
  \left[ - \frac{\mathcal{J}(\lbrace \bm{R}_{i} \rbrace,\lbrace
   \bm{A}_{j} \rbrace,\lbrace S_{j} \rbrace,Z)}{k_{B} T} \right] \\
   & = \frac{V}{\Lambda^{3}} \left(\frac{2 \pi b^{2}}{3 \Lambda^{2}}
   \right)^{3 (N - 1) / 2}
   \exp \left[ N e^{\nu / k_{B} T} \left(\frac{2 \pi N_{s} b^{2}}{3 \Lambda_{s}^{2}}
   \right)^{3 / 2} \right]
  \end{split}
\end{equation}
where $V$ is the system volume. $\Lambda$ and $\Lambda_{s}$ are
the thermal de Brogle wave lengths for a bead and a slip-spring,
respectively. (They are required to make the grand partition function
dimensionless. As we will show, thermodynamic properties are not
affected by them.)
The factorial $Z!$ is the Gibbs factor due to the indistinguishably of slip-springs.
The equilibrium distribution of a state is given as the following usual Boltzmann
form.
\begin{equation}
 \label{full_equilibrium_distribution_single_chain}
   P_{\text{eq}}(\lbrace \bm{R}_{i} \rbrace,\lbrace
   \bm{A}_{j} \rbrace,\lbrace S_{j} \rbrace,Z) = \frac{1}{\Xi} \frac{1}{\Lambda^{3 N} \Lambda_{s}^{3 Z} Z!}
   \exp \left[ - \frac{\mathcal{J}(\lbrace \bm{R}_{i} \rbrace,\lbrace
   \bm{A}_{j} \rbrace,\lbrace S_{j} \rbrace,Z)}{k_{B} T} \right]
\end{equation}
Here the equilibrium distribution function is normalized to satisfy the
following normalization condition.
\begin{equation}
 \label{full_equilibrium_distribution_single_chain_normalization_condition}
   \sum_{\lbrace S_{j} \rbrace, Z}
   \int d\lbrace \bm{R}_{i} \rbrace, d\lbrace \bm{A}_{j} \rbrace \,
   P_{\text{eq}}(\lbrace \bm{R}_{i} \rbrace,\lbrace
   \bm{A}_{j} \rbrace,\lbrace S_{j} \rbrace,Z) = 1
\end{equation}

The equilibrium average or distribution of a given set of variables can be
calculated from eq \eqref{full_equilibrium_distribution_single_chain}.
The equilibrium distribution of the chain conformation (the
bead positions $\lbrace \bm{R}_{i} \rbrace$), $P_{\text{eq}}(\lbrace
\bm{R}_{i} \rbrace)$, becomes
\begin{equation}
 \label{bead_position_distribution_single_chain}
  \begin{split}
   P_{\text{eq}}(\lbrace \bm{R}_{i} \rbrace)
   & = \sum_{\lbrace S_{j} \rbrace,Z}\int d\lbrace \bm{A}_{j} \rbrace \,
   P_{\text{eq}}(\lbrace \bm{R}_{i} \rbrace,\lbrace
   \bm{A}_{j} \rbrace,\lbrace S_{j} \rbrace,Z) \\
   & = \frac{1}{V} \left(\frac{3}{2 \pi b^{2}} \right)^{3 (N - 1) / 2}
   \exp \left[ - \sum_{i = 1}^{N - 1} \frac{3}{2 b^{2}} (\bm{R}_{i + 1} - \bm{R}_{i})^{2} \right]   
  \end{split}
\end{equation}
Eq \eqref{bead_position_distribution_single_chain} is the same as the
equilibrium conformation of an ideal chain. Thus we find that in
equilibrium our model can be reduced to a single ideal chain. Therefore all the
equilibrium statistical properties of a chain (such as the average end
to end vector or the average radius of gyration) are just the same as ones
of an ideal chain. This is consistent with the fact that the
entanglement effect is just a dynamic effect and does not affect static properties.

Similarly, we can calculate equilibrium statistical properties of
slip-springs.
The equilibrium distribution of the number of slip-springs, $P_{\text{eq}}(Z)$, is given as follows.
\begin{equation}
 \label{slip_spring_number_distribution_single_chain}
  \begin{split}
  P_{\text{eq}}(Z) & =
   \sum_{\lbrace S_{j} \rbrace} \int d\lbrace \bm{R}_{i} \rbrace d\lbrace
  \bm{A}_{j} \rbrace \, P_{\text{eq}}(\lbrace \bm{R}_{i} \rbrace,\lbrace
   \bm{A}_{j} \rbrace,\lbrace S_{j} \rbrace,Z) \\
   & = \frac{1}{Z!} \exp \left[ \frac{\tilde{\nu} Z}{k_{B} T} - e^{\tilde{\nu} / k_{B} T} \right]
  \end{split}
\end{equation}
where we defined the modified effective chemical potential $\tilde{\nu}$ as
\begin{equation}
 \label{modified_effective_chemical_potential_definition}
 \tilde{\nu} \equiv \nu
      + k_{B} T \ln \left[ N \left(\frac{2 \pi N_{s} b^{2}}{3 \Lambda_{s}^{2}}
   \right)^{3 / 2} \right]
\end{equation}
From eq \eqref{slip_spring_number_distribution_single_chain} we find
that the equilibrium distribution of $Z$ is expressed as a
Poisson distribution. This is consistent with the distribution
function in the slip-link model by Schieber \cite{Schieber-2003}.
The average number of slip-springs $\langle Z \rangle_{\text{eq}}$
($\langle \dots \rangle_{\text{eq}}$ means the equilibrium statistical average) can be related
to the modified effective chemical potential $\tilde{\nu}$.
\begin{equation}
 \label{average_slip_spring_number_single_chain}
 \langle Z \rangle_{\text{eq}} = \sum_{Z = 0}^{\infty} Z P_{\text{eq}}(Z) = e^{\tilde{\nu} / k_{B} T}
\end{equation}
We expect that the average number of slip-springs can
be expressed by using the characteristic number of beads between slip-springs,
$N_{0}$.
\begin{equation}
 \label{average_slip_spring_number_and_entanglement_bead_number_single_chain}
 \langle Z \rangle_{\text{eq}} = \frac{N}{N_{0}}
\end{equation}
Here we note that generally $N_{0}$ in eq
\eqref{average_slip_spring_number_and_entanglement_bead_number_single_chain}
does not coincide with the number of beads between entanglements
calculated from the plateau modulus, $N_{e}$. (Typically $N_{0}$ is
smaller than $N_{e}$
\cite{Masubuchi-Ianniruberto-Greco-Marrucci-2003,Likhtman-2005}.)
From eqs \eqref{average_slip_spring_number_single_chain} and
\eqref{average_slip_spring_number_and_entanglement_bead_number_single_chain},
we have the following relation between $N_{0}$ and $\tilde{\nu}$.
\begin{equation}
 \label{chemical_potential_and_entanglement_bead_number_single_chain}
 \tilde{\nu} = k_{B} T \ln \frac{N}{N_{0}}
\end{equation}
Eq \eqref{slip_spring_number_distribution_single_chain} can be then
rewritten simply as
\begin{equation}
 \label{slip_spring_number_distribution_single_chain_modified}
 P_{\text{eq}}(Z)
 = \frac{1}{Z!} \left(\frac{N}{N_{0}}\right)^{Z} e^{-N / N_{0}}
\end{equation}

By using eqs \eqref{bead_position_distribution_single_chain} and
\eqref{slip_spring_number_distribution_single_chain_modified}, we can rewrite eq \eqref{full_equilibrium_distribution_single_chain} as follows.
\begin{align}
 & \label{full_equilibrium_distribution_single_chain_modified}
  P_{\text{eq}}(\lbrace \bm{R}_{i} \rbrace,\lbrace
  \bm{A}_{j} \rbrace,\lbrace S_{j} \rbrace,Z)
   = P_{\text{eq}}(\lbrace \bm{R}_{i} \rbrace) P_{\text{eq}}(Z)
 \prod_{j = 1}^{Z} P_{\text{eq}}(S_{j}) P_{\text{eq}}(\bm{A}_{j} | \lbrace \bm{R}_{i}
 \rbrace, S_{j}) \\
 & \label{bead_index_distribution_single_chain}
 P_{\text{eq}}(S_{j}) \equiv \frac{1}{N} \\
 & \label{anchoring_point_distribution_single_chain}
 P_{\text{eq}}(\bm{A}_{j} | \lbrace \bm{R}_{i} \rbrace, S_{j}) \equiv
 \left(\frac{3}{2 \pi N_{s} b^{2}}\right)^{3 / 2} \exp \left[ 
 - \frac{3}{2 N_{s} b^{2}} (\bm{R}_{S_{j}} - \bm{A}_{j})^{2}  \right]
\end{align}
{ Here, the notation $P_{\text{eq}}(X | Y)$ represents the conditional probability of
$X$ for given $Y$.}
Eq \eqref{bead_index_distribution_single_chain} is the equilibrium
distribution of a bead index of a slip-spring. Eq
\eqref{bead_index_distribution_single_chain} can be interpreted as the
equilibrium distribution of an ideal gas particle on a one dimensional
lattice which has $N$ lattice points.
Eq \eqref{anchoring_point_distribution_single_chain} is the equilibrium
distribution of an anchoring point under a given chain conformation and
a bead index.

To study rheological properties, we need the microscopic expression for the stress tensor.
From the stress-optical rule, the stress tensor of a single chain can be
expressed by using the bead positions.
\begin{equation}
 \label{stress_tensor_single_chain}
 \bm{\sigma}(\lbrace \bm{R}_{i} \rbrace) = \sum_{i = 1}^{N - 1} \frac{3 k_{B} T}{b^{2}}
  (\bm{R}_{i + 1} - \bm{R}_{i}) (\bm{R}_{i + 1} - \bm{R}_{i}) - N k_{B} T \bm{1}
\end{equation}
The second term in the right hand side of eq \eqref{stress_tensor_single_chain}
does not have the non-diagonal elements and thus it can be dropped when
we consider the shear stress or the normal stress difference.
In equilibrium, the stress tensor of the system can be calculated to be
\begin{equation}
 \label{stress_tensor_single_chain_equilibrium_average}
  c_{0} \langle \bm{\sigma}(\lbrace \bm{R}_{i} \rbrace) \rangle_{\text{eq}} = - c_{0} k_{B} T \bm{1}
\end{equation}
where $c_{0}$ is the average density of polymer chains. ($c_{0}$ can be
related to the average bead density $\rho_{0}$ as $c_{0} = \rho_{0} /
N$). From eq \eqref{stress_tensor_single_chain_equilibrium_average} we
find that in equilibrium, our single chain slip-spring model gives the
stress tensor of an ideal gas system with the number density $c_{0}$.
This is physically natural because the entanglement effect is purely
kinetic, and the system is the same as a non-interacting
Rouse chain system in equilibrium.

\subsection{Dynamics}

We cannot study dynamical properties only from the grand potential \eqref{grand_potential_single_chain}.
In this subsection, we introduce a simple but physically valid dynamics
model for a single chain slip-spring model.
As we mentioned, the purpose of this work is to design a simulation
model which is suitable for the calculations on a GPU.
In this work, we will employ the CUDA programming model \cite{cuda-url,Lindholm-Nickolls-Oberman-Montrym-2008,nvidia-cuda-programming-guide}
which is developed and provided by NVIDIA corporation and widely used
for GPGPU calculations.

{ Although the CUDA programming model provides us a fast and
efficient GPGPU environment, there are several (rather strict) limitations.
Before we design the dynamics model, here we briefly review
some restrictions in CUDA which are directly related to the design of
our dynamics model.
First, parallel threads on a GPU are segmented into several blocks (each
block contains typically from several tens to several hundreds of
threads).
The data communication between different blocks are much slower compared
with the communication inside a block. Thus the data communication
between blocks should be reduced to achieve high performance.
Second, parallel threads are basically designed to perform the same task (with
different data values) and the complicated conditional branches decrease
the performance. Third, the amount of registers and shared memory is not
large (the total registers and shared memory are $8192$ ($32$kB) and $16$kB per
one block, respectively).
Fourth, some arithmetic operations are not implemented,
or not efficient compared with a CPU.
(Although these restrictions depend on the GPU
architecture, there are qualitatively similar restrictions for other
GPGPU calculation models.)
The dynamics model shown in this subsection is designed achieve high
performance calculations on a GPU under these limitations
(see Section \ref{implementaion_on_gpu}).}

Although we introduce a specific dynamics model in this
subsection, equilibrium properties (shown in the previous subsection)
are not altered for other dynamics models as long as the detailed
balance condition is satisfied.

We start from the dynamics of beads. In the absence of slip-springs ($Z =
0$), we expect that the dynamics reduces to the Rouse dynamics.
Then the dynamics of beads can be modelled as the simple Rouse type dynamics.
We use an overdamped Langevin equation as the dynamic equation for a bead.
\begin{equation}
 \label{langevin_equation_bead_position}
  \frac{d\bm{R}_{i}(t)}{dt} = - \frac{1}{\zeta} \frac{\partial \mathcal{J}(\lbrace \bm{R}_{i} \rbrace,\lbrace
   \bm{A}_{j} \rbrace,\lbrace S_{j} \rbrace,Z)}{\partial \bm{R}_{i}}
   + \bm{\kappa}(t) \cdot \bm{R}_{i} + \bm{\xi}_{i}(t)
\end{equation}
where $\zeta$ is the friction coefficient for a bead, $\bm{\kappa}(t)$
is the velocity gradient tensor, and $\bm{\xi}_{i}(t)$ is the Gaussian
white noise. $\bm{\xi}_{i}(t)$ satisfies the following
fluctuation-dissipation relation which guarantees the detailed balance.
\begin{align}
 & \langle \bm{\xi}_{i}(t) \rangle = 0 \\
 & \langle \bm{\xi}_{i}(t) \bm{\xi}_{j}(t') \rangle =
 \frac{2 k_{B} T}{\zeta} \delta_{ij} \delta(t - t') \bm{1}
\end{align}
where $\langle \dots \rangle$ means the statistical average and $\bm{1}$ is the unit tensor.

Dynamics of slip-springs is not trivial. Because we are seeking a model
which is suitable for simulations on a GPU, here we employ rather simple
dynamics for slip-springs.
We assume that the anchoring point of a slip-spring moves only by the external
flow. Thus the dynamic equation for $\bm{A}_{j}$ becomes the following
simple advection equation.
\begin{equation}
 \label{dynamic_equation_anchoring_position}
  \frac{d\bm{A}_{j}(t)}{dt} = \bm{\kappa}(t) \cdot \bm{A}_{j}
\end{equation}
In the absence of the velocity gradient, the anchoring
points do not change their positions unless they are reconstructed.
For the dynamics of slip-spring bead indices, we employ stochastic
jump processes \cite{vanKampen-book,Schieber-Neergaard-Gupta-2003}.
For simplicity, we assume that a slip-spring bead index can move only to its
neighboring bead indices by a single jump.
The jump probability from the bead index $S_{j}$ to the bead index $S_{j}'$ can be
expressed as a transition matrix $W(S_{j}' | S_{j})$.
%
%
%
\begin{equation}
 \label{transition_probability_s}
 W_{S}(S_{j}'|S_{j}) =
 \begin{cases}
  \displaystyle
  W_{S+}(S_{j}) & (S_{j}' = S_{j} + 1) \\
  W_{S-}(S_{j}) & (S_{j}' = S_{j} - 1) \\
  - W_{S+}(S_{j}) - W_{S-}(S_{j}) & (S_{j}' = S_{j}) \\
  0 & (\text{otherwise})
 \end{cases} 
\end{equation}
where $W_{S+}(S_{j})$ and $W_{S-}(S_{j})$ are transition probabilities which increment or
decrement the bead index, respectively. These transition probabilities should
satisfy the detailed balance condition. There are many possible forms for
the transition probabilities which satisfies the detailed balance
condition.
In this work we employ the following Glauber type dynamics \cite{Glauber-1963}. (It is  well
known that the Glauber dynamics satisfies the detailed balance condition.)
\begin{align}
 & \label{transition_probability_s_plus}
 W_{S+}(S_{j}) = 
 \begin{cases}
  \displaystyle
  \frac{k_{B} T}{\zeta_{s}} \left[ 1 - \tanh \left[ \frac{3}{4 N_{s} b^{2}}
  \left[(\bm{R}_{S_{j} + 1} - \bm{A}_{j})^{2} - (\bm{R}_{S_{j}} -
  \bm{A}_{j})^{2}\right] \right]
  \right] &
  { (S_{j} < N)} \\
  0 & { (S_{j} = N)}
 \end{cases} \\
 & \label{transition_probability_s_minus}
 W_{S-}(S_{j}) = 
 \begin{cases}
  \displaystyle
  \frac{k_{B} T}{\zeta_{s}} \left[ 1 - \tanh \left[ \frac{3}{4 N_{s} b^{2}}
  \left[(\bm{R}_{S_{j} - 1} - \bm{A}_{j})^{2} - (\bm{R}_{S_{j}} -
  \bm{A}_{j})^{2}\right] \right] \right] &
  { (S_{j} > 1)} \\
  0 & { (S_{j} = 1)}
 \end{cases}
\end{align}
Here $\zeta_{s}$ is a parameter which represents the effective friction
coefficient for a slip-spring. In contrast to the Likhtman's model, we
allow slip-springs to pass through each other. However, as mentioned by Likhtman, this
does not affect physical properties qualitatively. From the view point
of numerical calculations, this enables numerical schemes to be simple and
thus we can make the implementation on a GPU simple.

Finally we model the slip-spring reconstruction dynamics.
Slip-springs on chain ends ($S_{j} = 1,N$) can be removed from a chain
and destructed.
To compensate the destruction process, we have to introduce the
slip-spring construction process on chain ends. These reconstruction
events can be modelled as the jump processes, just like the dynamics for
$\lbrace S_{j} \rbrace$. If we assume that just one slip-spring can be
destructed or constructed at one reconstruction event, the jump probabilities
become as follows.
\begin{equation}
 \label{transition_probability_z}
 W_{Z}(Z'|Z) =
 \begin{cases}
  W_{Z+}(Z) & (Z' = Z + 1) \\
  W_{Z-}(Z) & (Z' = Z - 1) \\
  - W_{Z+}(Z) - W_{Z-}(Z) & (Z' = Z) \\
  0 & (\text{otherwise})
 \end{cases}
\end{equation}
where $W_{Z+}(Z)$ and $W_{Z-}(Z)$ are the construction and destruction
probabilities, respectively. From the detailed balance condition, they
should satisfy the following relation.
\begin{equation}
 \label{detail_balance_condition_transition_probability_z}
 W_{Z+}(Z) P_{\text{eq}}(\lbrace \bm{R}_{i} \rbrace,\lbrace \bm{A}_{j}
  \rbrace,\lbrace S_{j} \rbrace,Z)
  =
 W_{Z-}(Z) P_{\text{eq}}(\lbrace \bm{R}_{i} \rbrace,\lbrace \bm{A}_{j} \rbrace,\lbrace S_{j}
  \rbrace,Z + 1)
\end{equation}
Although the condition
\eqref{detail_balance_condition_transition_probability_z} limits the
form of dynamics, there are still many possible candidates for the reconstruction probabilities.
In this work, we employ the following rather simple jump probabilities for the slip-spring
reconstruction process.
\begin{align}
 & \label{transition_probability_z_plus}
  W_{Z+}(Z)  = \frac{k_{B} T}{\zeta_{s}}
 (\delta_{S_{Z + 1},1} +
 \delta_{S_{Z + 1},N}) \frac{1}{N_{0}} \left(\frac{3}{2 \pi N_{s} b^{2}}\right)^{3/2}
 \exp \left[ - \frac{3}{2 N_{s} b^{2}} (\bm{R}_{S_{Z + 1}} - \bm{A}_{Z +
 1})^{2} \right] \\
 & \label{transition_probability_z_minus}
 W_{Z-}(Z) = \frac{k_{B} T}{\zeta_{s}} \sum_{j = 1}^{Z} (\delta_{S_{j},1} +
 \delta_{S_{j},N})
\end{align}
It is straightforward to show that
eqs \eqref{transition_probability_z_plus} and
\eqref{transition_probability_z_minus} togather with eq
\eqref{full_equilibrium_distribution_single_chain_modified} satisfy the
condition \eqref{detail_balance_condition_transition_probability_z}.
As we will show later, the transition probabilities
\eqref{transition_probability_z_plus} and
\eqref{transition_probability_z_minus} enable simple numerical
implementations which are suitable to simulations on a GPU.

The single chain slip-spring model shown in this subsection can
reproduce reptation type dynamics \cite{Doi-Edwards-book} qualitatively.
We emphasize that each dynamics described in this subsection satisfies
the detailed
balance condition, and thus our single chain slip-spring model has
well-defined thermodynamic equilibrium state.

Although
we did not incorporate the effects such as the constraint release (CR) or
the convective constraint release (CCR) into our model, it is not
difficult to take these effects and improve the model.
Interestingly, while the CR process is not explicitly considered,
the relaxation process which is qualitatively similar to the CR exists
in our model. (As shown in Appendix \ref{constraint_release_type_effect},
the dynamics of bead indices exhibit CR type relaxation mechanism.)
Our model can reproduce rheological properties reasonably, as we will
show later.

\subsection{Linear Response Theory}
\label{linear_response_theory}

To obtain linear response functions such as the shear relaxation modulus
around equilibrium, the linear response
theory \cite{Evans-Morris-book} is sometimes quite useful.
{ The linear response theory states that if the dynamics
satisfies the detailed balance condition, the response of the system to a small
perturbation is simply expressed by correlation functions (the
fluctuation-dissipation relation). Because our model is designed to
satisfy the detailed balance condition, we can utilize the
fluctuation-dissipation relation to calculate the response in our model.
(Some slip-link models do not satisfy the detailed balance
condition. For such models, the validity of the linear response theory
is not guaranteed and it may not be useful.)}
In this subsection, we show the explicit expression for the
relaxation modulus by using the standard linear response theory.

The dynamics of our single chain slip-spring model is described by the
Langevin equation and jump processes. In the absence of the velocity
gradient tensor the system can relax to the equilibrium state, because
the detailed balance condition is satisfied. Now we consider the velocity
gradient tensor, $\bm{\kappa}(t)$, as a time-dependent small
perturbation and calculate the response of the time-dependent stress
tensor $\bm{\sigma}(t)$ to $\bm{\kappa}(t)$.

The probability distribution function is useful to calculate the linear response.
We define the time-dependent probability distribution function in the phase space as
\begin{equation}
 \label{probability_distribution_single_chain}
 P(\lbrace \bm{r}_{i} \rbrace,\lbrace
 \bm{a}_{i} \rbrace,\lbrace s_{i} \rbrace,z;t)
 \equiv \left\langle \delta_{z,Z(t)}
 \left[ \prod_{i = 1}^{N} \delta(\bm{r}_{i} - \bm{R}_{i}(t))
 \right] \Bigg[ \prod_{j = 1}^{z}
 \delta_{s_{j},S_{j}(t)} \delta(\bm{a}_{j} - \bm{A}_{j}(t)) \Bigg]
 \right\rangle
\end{equation}
The time evolution of the probability distribution function can be given
as the following master equation.
\begin{equation}
 \label{time_evolution_probability_distribution_single_chain}
 \frac{\partial}{\partial t} P(\lbrace \bm{r}_{i} \rbrace,\lbrace
 \bm{a}_{j} \rbrace,\lbrace s_{j} \rbrace,z;t)
 = \left[ {\mathcal{L}}_{\bm{R}}(t)
    + {\mathcal{L}}_{\bm{A}}(t)
    + {\mathcal{L}}_{S}
    + {\mathcal{L}}_{Z}
   \right]
  P(\lbrace \bm{r}_{i} \rbrace,\lbrace
 \bm{a}_{j} \rbrace,\lbrace s_{j} \rbrace,z;t)
\end{equation}
where ${\mathcal{L}}_{\bm{R}}(t)$ and ${\mathcal{L}}_{\bm{A}}(t)$ are the
Fokker-Planck type operators, and
${\mathcal{L}}_{S}$ and ${\mathcal{L}}_{Z}$ are the time evolution
operators which come from jump processes for $\lbrace S_{j} \rbrace$ and $Z$.
Notice that ${\mathcal{L}}_{\bm{R}}(t)$ and ${\mathcal{L}}_{\bm{A}}(t)$
depend on time $t$ whereas ${\mathcal{L}}_{S}$ and ${\mathcal{L}}_{Z}$
do not. This is because the dynamics of $\lbrace S_{j} \rbrace$ and $Z$
is not directly affected by the external flow.
The explicit forms for ${\mathcal{L}}_{\bm{R}}(t)$ and
${\mathcal{L}}_{\bm{A}}(t)$ are as follows.
\begin{equation}
 {\mathcal{L}}_{\bm{R}}(t) P
 = \sum_{i = 1}^{N} \frac{1}{\zeta} \frac{\partial}{\partial \bm{r}_{i}} \cdot
 \left[ \frac{\partial \mathcal{J}(\lbrace \bm{r}_{i} \rbrace,\lbrace
  \bm{a}_{j} \rbrace, \lbrace s_{j} \rbrace,z)}{\partial
 \bm{r}_{i}} P
 - \zeta \bm{\kappa}(t) \cdot \bm{r}_{i} P + k_{B} T
 \frac{\partial P}{\partial \bm{r}_{i}} \right]
\end{equation}
\begin{equation}
 {\mathcal{L}}_{\bm{A}}(t) P
 = - \sum_{j = 1}^{Z} \frac{\partial}{\partial \bm{a}_{j}} \cdot
 \left[ \bm{\kappa}(t) \cdot \bm{a}_{j} P \right]
\end{equation}
We do not show the explicit forms for ${\mathcal{L}}_{S}$ and
${\mathcal{L}}_{Z}$ here, because their explicit forms are not
required in the following calculations.

To calculate the linear response, we need to decompose the time
evolution operator into the equilibrium and perturbation parts.
Thus we define the following two time evolution operators.
\begin{align}
 & {\mathcal{L}}_{0} P \equiv
 \sum_{i = 1}^{N} \frac{1}{\zeta} \frac{\partial}{\partial \bm{r}_{i}} \cdot
 \left[ \frac{\partial \mathcal{J}}{\partial
 \bm{r}_{i}} P + k_{B} T
 \frac{\partial P}{\partial \bm{r}_{i}} \right]
    + {\mathcal{L}}_{S} P
    + {\mathcal{L}}_{Z} P \\
 & {\mathcal{L}}_{1}(t) P \equiv - \sum_{i = 1}^{N} 
 \frac{\partial}{\partial \bm{r}_{i}} \cdot
 \left[ \bm{\kappa}(t) \cdot \bm{r}_{i} P \right] - \sum_{j = 1}^{Z} \frac{\partial}{\partial \bm{a}_{j}} \cdot
 \left[ \bm{\kappa}(t) \cdot \bm{a}_{j} P \right]
\end{align}
By using these operators, the time evolution equation of the probability distribution function can
be simply described as follows.
\begin{equation}
 \label{time_evolution_probability_distribution_single_chain_modified}
 \frac{\partial P}{\partial t} = [{\mathcal{L}}_{0} + {\mathcal{L}}_{1}(t)] P
\end{equation}
In the absence of the velocity gradient, the perturbation part of the
time evolution operator disappears ($\mathcal{L}_{1}(t) = 0$) and
the system relaxes to the equilibrium state. The equilibrium
distribution function $P_{\text{eq}}$ is given as the Boltzmann form
since the detailed balance condition is satisfied.
\begin{equation}
 P_{\text{eq}}(\lbrace \bm{r}_{i} \rbrace,\lbrace
 \bm{a}_{i} \rbrace,\lbrace s_{i} \rbrace,z) \equiv
 \frac{1}{\Xi} \frac{1}{\Lambda^{3 N} \Lambda_{s}^{3 z} z!} \exp \left[-\frac{\mathcal{J}(\lbrace \bm{r}_{i} \rbrace,\lbrace
 \bm{a}_{i} \rbrace,\lbrace s_{i} \rbrace,z)}{k_{B} T}\right]
\end{equation}
Using the equilibrium distribution function we write the equilibrium average for an operator $\hat{B}$
as
\begin{equation}
 \langle \hat{B} \rangle_{\text{eq}} \equiv \sum_{z,\lbrace s_{j} \rbrace}
  \int d\lbrace \bm{r}_{i} \rbrace d\lbrace \bm{a}_{j} \rbrace \,
 \hat{B} P_{\text{eq}}(\lbrace \bm{r}_{i} \rbrace,\lbrace
 \bm{a}_{i} \rbrace,\lbrace s_{i} \rbrace,z)
\end{equation}

Now we can follow the standard procedure to calculate the linear
response \cite{Evans-Morris-book,Risken-book}. The time-dependent stress tensor of a single chain is calculated by the
following stress tensor operator.
\begin{equation}
 \label{stress_tensor_operator_single_chain}
 \hat{\bm{\sigma}}
 \equiv \sum_{i = 1}^{N - 1}
  \frac{3 k_{B} T}{b^{2}} (\bm{r}_{i + 1} -
  \bm{r}_{i}) (\bm{r}_{i + 1} -
  \bm{r}_{i}) - N k_{B} T \bm{1}
\end{equation}
Following the standard linear response theory, we have the following
expression for the time-dependent stress tensor. (See Appendix \ref{detail_calculations_in_linear_response_theory}
for detail.)
\begin{equation}
 \begin{split}
 \bm{\sigma}(t)
  & = \sum_{z,\lbrace s_{j} \rbrace} \int d\lbrace \bm{r}_{i} \rbrace
  d\lbrace \bm{a}_{j} \rbrace \, \hat{\bm{\sigma}} P(\lbrace \bm{r}_{i} \rbrace, \lbrace
  \bm{a}_{j} \rbrace,\lbrace s_{j} \rbrace, z; t) \\
  & = \bm{\sigma}_{\text{eq}} +
  \int_{-\infty}^{t} dt'
  \sum_{z,\lbrace s_{i} \rbrace} \int d\lbrace \bm{r}_{i} \rbrace
  d\lbrace \bm{a}_{i} \rbrace \, \hat{\bm{\sigma}}
  e^{(t - t'){\mathcal{L}}_{0}} {\mathcal{L}}_{1}(t') P_{\text{eq}}  
 \end{split}
\end{equation}
where $\bm{\sigma}_{\text{eq}} \equiv \langle \hat{\bm{\sigma}}
\rangle_{\text{eq}} = - k_{B} T \bm{1}$ is the equilibrium value of the stress tensor.
After straightforward calculations, 
finally we have the following expression for the time-dependent stress
tensor.
\begin{equation}
 \label{single_chain_time_depending_stress_expression_final}
 \begin{split}
 \bm{\sigma}(t)
  &  = \bm{\sigma}_{\text{eq}} + \frac{1}{k_{B} T} \int_{-\infty}^{t} dt' \,
  \left[ \langle \hat{\bm{\sigma}}(t - t')
  \hat{\bm{\sigma}} \rangle_{\text{eq}}
  + \langle \hat{\bm{\sigma}}(t - t') \hat{\bm{\sigma}}^{(v)} \rangle_{\text{eq}}
  \right] : \bm{\kappa}(t') \\
 \end{split}
\end{equation}
where $\hat{\bm{\sigma}}(t)$ is the stress tensor operator evolved by time
$t$ (the time-shifted stress tensor operator) and
$\hat{\bm{\sigma}}^{(v)}$ is the virtual stress tensor operator
by slip-springs and $:$ means the dyadic product. $\hat{\bm{\sigma}}^{(v)}$ is defined as
\begin{equation}
 \label{virtual_stress_tensor_operator_single_chain}
 \hat{\bm{\sigma}}^{(v)} \equiv
  \sum_{j = 1}^{z - 1} \frac{3 k_{B} T}{N_{s} b^{2}} (\bm{r}_{s_{j}} -
  \bm{a}_{j})
  (\bm{r}_{s_{j}} - \bm{a}_{j}) -  z k_{B} T \bm{1}
\end{equation}
Eq \eqref{virtual_stress_tensor_operator_single_chain} is similar to
eq \eqref{stress_tensor_operator_single_chain} but it represents the
contribution of slip-springs.
%
The relaxation modulus tensor $\bm{G}(t)$ (which is a
fourth order tensor) corresponds to the response function of the stress
to the velocity gradient. In this work we use the following
definition for the relaxation modulus tensor.
\begin{equation}
 \label{relaxation_modulus_tensor_definition}
 c_{0} [\bm{\sigma}(t) - \bm{\sigma}_{\text{eq}}]
  = \int_{-\infty}^{t} dt' \,
  \bm{G}(t - t') : \bm{\kappa}(t') \\
\end{equation}
From eqs \eqref{single_chain_time_depending_stress_expression_final} and
\eqref{relaxation_modulus_tensor_definition},
we find that the relaxation modulus tensor is simply given as
\begin{equation}
 \label{relaxation_modulus_tensor_green_kubo_form}
 \bm{G}(t) = \frac{c_{0}}{k_{B} T} \langle \hat{\bm{\sigma}}(t)
  \hat{\bm{\sigma}} \rangle_{\text{eq}}
  + \frac{c_{0}}{k_{B} T} \langle \hat{\bm{\sigma}}(t) \hat{\bm{\sigma}}^{(v)} \rangle_{\text{eq}}
\end{equation}
It should be noted here that there are two contributions for the
relaxation modulus tensor.
One is the usual stress-stress autocorrelation function, and
another is the stress-virtual stress correlation function which does not
appear in usual many body systems.

Here we note that eq \eqref{relaxation_modulus_tensor_green_kubo_form}
coincide with the formula previously proposed by Ramirez, Sukumaran
and Likhtman \cite{Ramirez-Sukumaran-Likhtman-2007}.
Intuitively, this form can be understood as follows.
The stress tensor operator
$\hat{\bm{\sigma}}$ is not conjugate to the perturbation
$\bm{\kappa}(t)$, but the sum of the stress and the virtual stress operators,
$\hat{\bm{\sigma}} + \hat{\bm{\sigma}}^{(v)}$, is conjugate to the
perturbation. Then, following the standard formula of
the linear response theory, the response function is given as the time
correlation function of $\hat{\bm{\sigma}}$ and $\hat{\bm{\sigma}} +
\hat{\bm{\sigma}}^{(v)}$, which gives eq \eqref{single_chain_time_depending_stress_expression_final}.
If we employ $\hat{\bm{\sigma}} + \hat{\bm{\sigma}}^{(v)}$ as the stress
tensor of a chain, the relaxation modulus tensor is simply given as the
autocorrelation function of
$\hat{\bm{\sigma}} + \hat{\bm{\sigma}}^{(v)}$. However, such a
definition violates the stress-optical rule, and it is physically
unnatural.
{ (We will examine the contribution of the
virtual stress is quantitatively, later.)}
Therefore, in
this work we use $\hat{\bm{\sigma}}$ as the stress tensor operator.

\section{Numerical Scheme and Implementation}

\subsection{Discretization Scheme}

To solve dynamics of our single chain slip-spring model numerically, we need to
discretize dynamic equations and jump processes. In this subsection, we
briefly show the discretization scheme. Since our purpose in this work
is to develop a model which is suitable for calculations on a GPU, here
we aim to make rather simple and stable schemes.

Before considering discretization schemes, we make all the parameters
dimensionless. We set $b = 1$, $k_{B} T = 1$, and $\zeta = 1$. (This is
equivalent to make all the dimensional parameters dimensionless by
characteristic scales $b$, $k_{B} T$, and $\zeta$.) The characteristic
time scale of simulations also becomes unity ($\tau_{0} \equiv \zeta b^{2} / k_{B} T = 1$).

To make schemes simpler, we split dynamics from $t$ to $t + \Delta t$
(with $\Delta t$ being the time step size) into several substeps.
In this work, we use the following three substeps to evolve the system
from $t$ to $t + \Delta t$. Each substep is designed to satisfy the
detailed balance condition for small $\Delta t$ ($\Delta t \ll 1$).
\begin{enumerate}
 \item Integrate dynamics equations for $\lbrace \bm{R}_{i} \rbrace,
       \lbrace \bm{A}_{j} \rbrace$
       (eqs \eqref{langevin_equation_bead_position} and
       \eqref{dynamic_equation_anchoring_position}).
       These are stochastic differential equation and ordinary
       differential equation and thus we can employ several standard
       schemes.
       To minimize the computational cost, here we employ the explicit
       Euler method which is not accurate but the simplest.
\begin{align}
 & \bm{R}_{i}(t + \Delta t)
 = - \Delta t \frac{\partial \mathcal{J}(\lbrace \bm{R}_{i} \rbrace,\lbrace
   \bm{A}_{j} \rbrace,\lbrace S_{i} \rbrace,Z)}{\partial \bm{R}_{i}}
   + \Delta t \bm{\kappa}(t) \cdot \bm{R}_{i} + \sqrt{2 \Delta t} \bm{w}_{i} \\
 &  \bm{A}_{j}(t + \Delta t) = \Delta t \bm{\kappa}(t) \cdot \bm{A}_{j}
\end{align}
       where $\bm{w}_{i}$ is a standard distribution random number vector which satisfies
       the following relation.
\begin{align}
 & \langle \bm{w}_{i} \rangle = 0 \\
 & \langle \bm{w}_{i} \bm{w}_{j} \rangle =
 \delta_{ij} \bm{1}
\end{align}
       $\bm{w}_{i}$ can be generated easily by some random number
       generators such as the combination of the linear coagulation
       method and the Box-Muller transform.
 \item Move slip-spring bead indices $\lbrace S_{j} \rbrace$ by transition probabilities
       \eqref{transition_probability_s}-\eqref{transition_probability_s_minus}.
       We use the following accumulated probabilities for finite time step
       $\Delta t$ and uniform distribution random variables.
\begin{equation}
 P(S_{j} \to S_{j} + 1) = 
 \begin{cases}
  \begin{aligned}
   & \displaystyle
   \min \bigg\lbrace   \frac{\Delta t}{\zeta_{s}} \bigg[ 1 - \tanh \bigg[ \frac{3}{4 N_{s}}
  \big[(\bm{R}_{S_{j} + 1} - \bm{A}_{j})^{2} \\
   & \qquad \qquad - (\bm{R}_{S_{j}} -
  \bm{A}_{j})^{2}\big] \bigg] \bigg] ,
   \frac{1}{2} \bigg\rbrace
  \end{aligned} &
  { (S_{j} < N)} \\
  0 & { (S_{j} = N)}
 \end{cases}
\end{equation}
\begin{equation}
 P(S_{j} \to S_{j} - 1) = 
 \begin{cases}
  \begin{aligned}
   & \displaystyle
   \min \bigg\lbrace   \frac{\Delta t}{\zeta_{s}} \bigg[ 1 - \tanh \bigg[ \frac{3}{4 N_{s}}
  \big[(\bm{R}_{S_{j} - 1} - \bm{A}_{j})^{2} \\
   & \qquad \qquad - (\bm{R}_{S_{j}} -
  \bm{A}_{j})^{2}\big] \bigg] \bigg] ,
   \frac{1}{2} \bigg\rbrace
  \end{aligned} &
  { (S_{j} > 1)} \\
  0 & { (S_{j} = 1)}
 \end{cases}
\end{equation}
The minima are taken so that each transition probability does not
exceed $1/2$. Although the discretization scheme shown here is not accurate,
       it does not fail even for large $\Delta t$ due to this trick.
 \item Reconstruct slip-springs on chain ends by the transition probability
       eq \eqref{transition_probability_z},
       \eqref{transition_probability_z_plus}, and \eqref{transition_probability_z_minus}.
       First, new slip-springs are constructed at chain ends by the
       following accumulated probability. The construction is
       attempted $K$ times repeatedly (with $K$ being an integer parameter which can depend
       on $Z$). { This parameter $K$ is introduced to allow
       the construction attempt evaluated in parallel. This is because
       parallel attempts are much efficient than single attempt on a
       GPU. (See Section
       \ref{implementaion_on_gpu}.)}
\begin{align}
 & P_{\text{construct}}(S_{Z + 1} = 1) = \min \left\lbrace \frac{\Delta t}{\zeta_{s}}
 \frac{1}{N_{0}} \frac{1}{K}, \frac{1}{2} \right\rbrace \\
 & P_{\text{construct}}(S_{Z + 1} = N) = \min \left\lbrace \frac{\Delta t}{\zeta_{s}}
 \frac{1}{N_{0}} \frac{1}{K}, \frac{1}{2} \right\rbrace
\end{align}
       The minima are taken to prevent each transition probability from
       exceeding $1/2$.
       The anchored point
       position of a new slip-spring, $\bm{A}_{Z + 1}$, is sampled from the equilibrium
       distribution. Since the equilibrium distribution of an anchoring
       point is Gaussian (eq
       \eqref{anchoring_point_distribution_single_chain}), it can be easily generated.
\begin{equation}
 P_{\text{construct}}(\bm{A}_{Z + 1})
  = \left(\frac{3}{2 \pi N_{s}}\right)^{3/2}
  \exp \left[ - \frac{3}{2 N_{s}} (\bm{R}_{S_{Z + 1}}
	- \bm{A}_{Z + 1})^{2}\right]
\end{equation}
       Second, each slip-spring is destructed by the following
       accumulated probability.
\begin{equation}
 P_{\text{destruct}}(S_{j}) = \min \left\lbrace \frac{\Delta t}{\zeta_{s}} (\delta_{S_{j},1} +
 \delta_{S_{j},N}), 1 \right\rbrace \\ 
\end{equation}
       As before, the minimum is taken to avoid each transition probability to
       exceed $1$.
       We reset indices of slip-springs
       after the constructions or destructions. This ensures that the
       index for slip-springs on a chain
       always runs from $1$ to $Z$ ($j = 1,2,\dots,Z$).
\end{enumerate}

The discretization schemes shown here are not accurate compared with
more advanced schemes.
(For example, it is possible to improve the
accuracy by
employing more advanced and accurate discretization schemes
such as the stochastic Runge-Kutta method \cite{Honeycutt-1992}.)
In this work, we prefer inaccurate but simple schemes for an
implementation on a GPU. Since discretization schemes are designed to be
as simple as possible, it is not difficult to implement them for GPGPU
calculations (in which we have several limitations due to the GPU architecture).
Although the accuracy is not high, we can perform stable simulations with
these schemes even for not small $\Delta t$. (That is, the simulations do
not fail easily even for rather large $\Delta t$.)

Before starting simulations we have to prepare well equilibrated
samples. Since we know the equilibrium probability distribution, we can
directly generate the equilibrium conformation of chains and
slip-springs easily. All the simulations performed in this work start
from the equilibrium state generated based on eq
\eqref{full_equilibrium_distribution_single_chain_modified}. The sampling
procedure is as follows.
\begin{enumerate}
 \item Generate polymer conformation $\lbrace \bm{R}_{i} \rbrace$ by
       sampling from the Gaussian distribution (eq \eqref{bead_position_distribution_single_chain}).
 \item Sample $Z$ from the Poisson distribution (eq \eqref{slip_spring_number_distribution_single_chain_modified}).
 \item Generate $Z$ slip-springs. Each segment index, $S_{j}$, is sampled
       from the uniform distribution (eq \eqref{bead_index_distribution_single_chain}) and each
       anchoring point, $\bm{A}_{j}$, is sampled from the Gaussian
       distribution (eq \eqref{anchoring_point_distribution_single_chain}).
\end{enumerate}

To obtain statistical quantities form Langevin type simulations,
we need to perform many simulations with the same parameter set and
different random number series. For this purpose, we simulate $M$
different polymer chains and calculate the statistical average of an
arbitrary physical quantity $B$ approximately as
\begin{equation}
 \langle B \rangle
  \approx \frac{1}{M} \sum_{k = 1}^{M} B(\lbrace \bm{R}_{k,i} \rbrace,\lbrace \bm{A}_{k,j} \rbrace,
  \lbrace S_{k,j} \rbrace,Z_{k})
\end{equation}
where the subscript $k$ means that the variable is of the $k$-th
sample chain (for example, $\bm{R}_{k,i}$ is the $i$-th bead position of the
$k$-th chain).

\subsection{Implementation on GPU}
\label{implementaion_on_gpu}

As we mentioned, we use the CUDA programming model to implement the
discretized single chain slip-spring model on a GPU.
In the CUDA
programming model, there are two memory spaces on a graphics card; the
global memory and the shared memory. The shared memory is small
(currently its size is just $16\text{kB}$) but the access speed is fast
(comparable to the access speed of the registers). The global memory is
large but the access speed is rather slow. Thus it is required to reduce
the access to the global memory to speed up the program.
Fortunately, our single chain
slip-spring model requires very small memories and thus most of the data can
be stored in the registers or the shared memory.
To make the CUDA program efficient, it is also required to achieve high
parallelism because a GPU has many threads (typically about several
hundreds or several thousands) which are executed simultaneously.
We can implement highly parallelized program by allocating one bead and
one slip-spring to one CUDA thread.
Although this limits the maximum number of
slip-springs ($Z \le N$), in most cases it cause no problems
(the probability that $Z > N$ is practically
negligible if $N_{0}$ is not small). Here we note that, it is difficult to implement in such a way if we
employ the slip-link type singe chain model
\cite{Hua-Schieber-1998,Masubuchi-Takimoto-Koyama-Ianniruberto-Greco-Marrucci-2001,Doi-Takimoto-2003,Schieber-Neergaard-Gupta-2003,Nair-Schieber-2006},
because nodes are dynamically
reconstructed and thus the number of total nodes is not constant.
This is one of the main reason why the slip-spring model is employed in
this work.

For the slip-spring reconstruction scheme on a GPU, we set the number of
attempts
to construct slip-springs as $K = N - Z$. This means that each
thread which do not have a slip-spring attempts to construct a new
one (and each attempt can be evaluated in parallel). For a CPU, we set
$K = 1$, which means that the slip-spring construction is attempted only
once at each step. Clearly this is suitable for a CPU, because we cannot
evaluate many attempts in parallel.
(Strictly speaking, this implementation will violate
the detailed balance condition. But the violation is practically
negligible.)

Currently it is generally not efficient to use double precision floating-point
numbers (``double'' type variables in CUDA), and thus single precision
floating-point numbers (the ``float'' type variables) should be used to
accelerate simulations efficiently. Fortunately this limitation is not serious for
our single chain slip-spring model. Because our model is based on the
stochastic differential equations and jump processes, the accuracy is
mainly determined by number of samples and the error is typically much
larger than the error of single precision floating-point number operations
(the machine epsilon is about $\epsilon_{\text{float}} \approx 10^{-7}$
whereas the statistical error is roughly proportional to the inverse
square root of sampling numbers).
Several elementary mathematical functions can be calculated very
efficiently on a GPU. In our implementation, functions such as $\sin x$,
$\cos x$, $\exp x$, $\ln x$, or $\sqrt{x}$ are used. Although the fast
calculations for these mathematical functions
involve some errors, such errors are not serious in our simulations (the errors
are typically comparable to one of single precision floating-point
number operations).

In parallelized programs, it is often needed to perform reduction
operations. In our model, we have to calculate the total number of
slip-springs on a chain or forces acting on beads caused by
slip-springs. The reduction operations are one of the most
time-consuming parts in our simulations.
Although CUDA currently does not provide special functions for these
reduction operations, we can use some techniques to improve the
speed of the reduction operations \cite{cuda-sdk-reduction}.

Another time-consuming part is the calculation of the slip-spring force
acting on a chain. { To calculate the slip-spring force efficiently, we
utilize the fixed-point real number technique
\cite{Narumi-Sakamaki-Kameoka-Yasuoka-2008} and the atomic operations
\cite{nvidia-cuda-programming-guide} in CUDA.
The fixed-point real number technique uses the integer type variable $x$
as the real number $y = x \times \epsilon_{\text{fixed}}$ with
$\epsilon_{\text{fixed}}$ being the resolution. The resolution $\epsilon_{\text{fixed}}$ should be
determined so that the truncation error is sufficiently small and the
overflow does not occur. 
(Because the single precision float number has
the machine epsilon $\epsilon_{\text{float}} \approx 10^{-7}$, $\epsilon_{\text{fixed}}$ is not necessarily to be very small.)
In this work we set $\epsilon_{\text{fixed}} = 1 / 4096 \approx 2.4
\times 10^{-4}$, which has a sufficient resolution yet the overflow does not
occur even under fast shear rates.
The results shown below are not
sensitive to the value of $\epsilon_{\text{fixed}}$, as long as
$\epsilon_{\text{fixed}}$ is not too small nor too large.}


\section{Results}

\subsection{Rheological Properties}

Before measuring the acceleration effect by a GPU,
we calculate several rheological properties of our slip-spring
model. To obtain reliable simulation results, in this subsection, all
the simulations are performed on a CPU. (The discretization schemes on a
CPU are almost the same as the schemes on a GPU.)
We set the simulation parameters as follows; the entanglement bead number $N_{0} = 4$,
slip-spring strength $N_{s} = 0.5$, the slip-spring friction coefficient
$\zeta_{s} = 0.1$, and the time step size $\Delta t = 0.01$.

\begin{figure}[htb]
 \centering
 {\includegraphics[width=0.95\linewidth,clip]{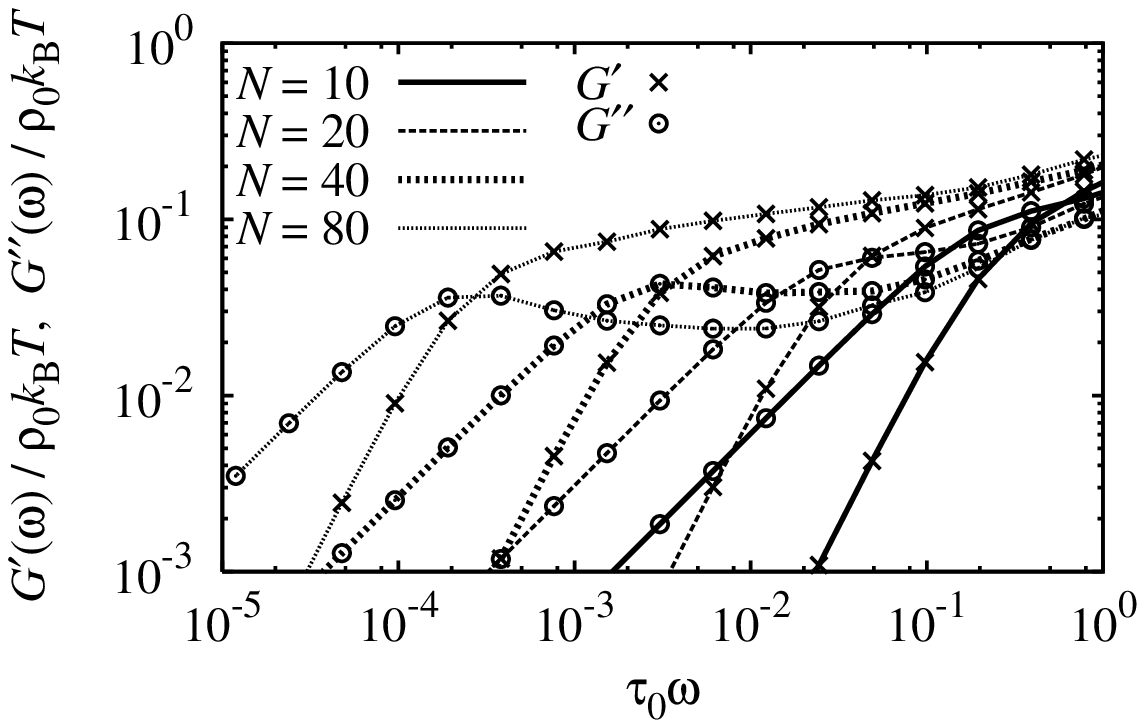}}
 \caption{\label{storage_and_loss_moduli}}
\end{figure}

We first calculate the linear viscoelasticity by using the linear
response formula \eqref{relaxation_modulus_tensor_green_kubo_form}.
The storage and loss moduli, $G'(\omega)$ and $G''(\omega)$, are then
calculated by utilizing the following relations.
\begin{align}
 & \label{storage_modulus_definition}
 G'(\omega) = \omega \int_{0}^{\infty} dt \, G_{xyxy}(t) \sin(\omega t) \\
 & \label{loss_modulus_definition}
 G''(\omega) = \omega \int_{0}^{\infty} dt \, G_{xyxy}(t) \cos(\omega t)
\end{align}
The integration in eqs \eqref{storage_modulus_definition} and
\eqref{loss_modulus_definition} are numerically calculated by using the
trapezoidal rule.
Figure \ref{storage_and_loss_moduli} shows the linear viscoelasticity data calculated from equilibrium
single chain slip-springs simulations (for $N = 10,$ $20,$ $40,$ and
$80$).
Notice that $G'(\omega)$ and $G''(\omega)$ are normalized by $\rho_{0} k_{B} T$ to
be dimensionless forms ($\rho_{0}$ is the average bead density).
As shown in Figure \ref{storage_and_loss_moduli},
our model reproduces linear viscoelasticity of
entangled polymers qualitatively well, especially considering the
simplicity of the model.
The plateau modulus is about $G_{N}^{(0)} \approx 0.1 \rho_{0} k_{B} T$,
which is almost the same as the plateau modulus obtained by the original
slip-spring model simulation \cite{Likhtman-2005}. This means that, the
characteristic number of beads between slip-springs $N_{0}$ in this model is not identical to the
entanglement bead number calculated from the plateau modulus based on
the rubber elasticity theory. Following the standard definition, we
define $N_{e}$ via the following equation.
\begin{equation}
 \label{entanglement_bead_number_rubber_elasticity_definition}
 N_{e} \equiv \frac{\rho_{0} k_{B} T}{G_{N}^{(0)}}
\end{equation}
From $G_{N}^{(0)} \approx 0.1 \rho_{0} k_{B} T$, we have $N_{e} \approx
10$. This value is much larger than $N_{0}$ ($N_{e} \approx 2.5 N_{0}$).
Therefore the plateau modulus or $N_{e}$ are not
determined only by $N_{0}$, but they depends on various parameters in
a rather complex way. Practically, it is reasonable to use both $N_{0}$
and $G_{0}$ (or $G_{N}^{(0)}$) as two independent fitting parameters \cite{Likhtman-2005,Masubuchi-Ianniruberto-Greco-Marrucci-2008}.

\begin{figure}[htb]
 \centering
 {\includegraphics[width=0.95\linewidth,clip]{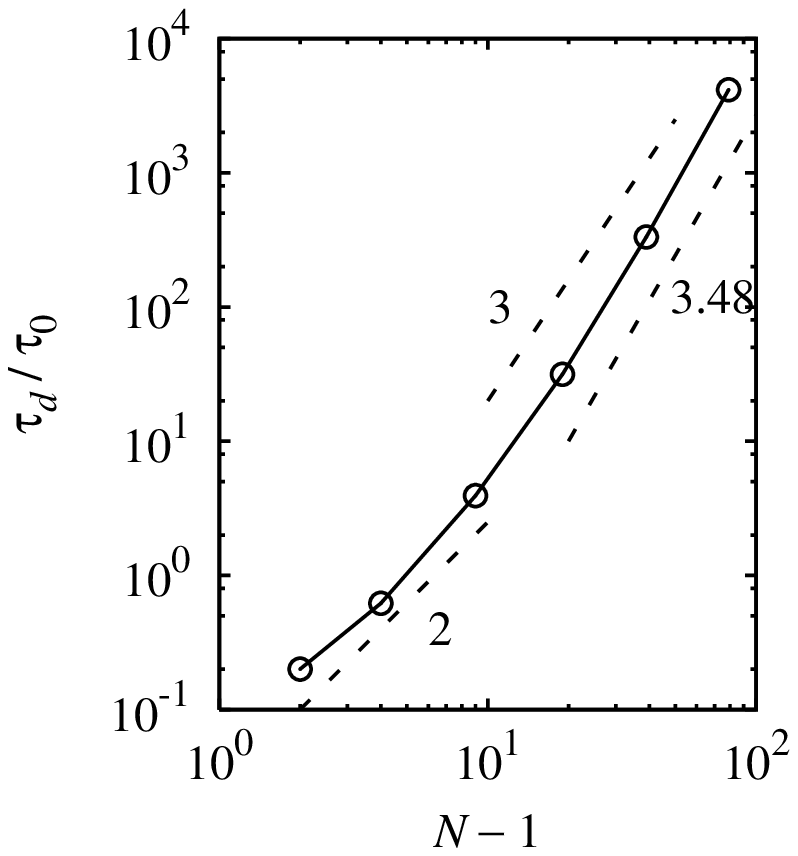}}
 \caption{\label{longest_relaxation_time}}
\end{figure}

Figure \ref{longest_relaxation_time} shows the longest relaxation time
calculated from the linear viscoelasticity data.
The longest relaxation time is calculated via the following equation
\begin{equation}
 \frac{1}{\tau_{d}} = - \lim_{t \to \infty} \frac{\ln G_{xyxy}(t)}{t}
\end{equation}
For small $N$, we find that the
longest relaxation time is proportional to $(N - 1)^{2}$, which agrees
with the scaling of the Rouse relaxation time. (Here $N - 1$ is used
instead of $N$, since $N - 1$ corresponds to the number of bonds in a
chain \cite{Strobl-book}.)
This indicates that for
small $N$ (from Figure \ref{longest_relaxation_time}, $N \lesssim 10$) a
chain with slip-springs behave essentially as an unentalgled chain.
For large $N$, the relaxation time depends on $N - 1$ as
$\tau_{d} \propto (N - 1)^{3.48}$,
which is
similar to experimental results, $\tau_{d} \propto (N - 1)^{3.4}$
(larger than the prediction of the pure reptation theory \cite{Doi-Edwards-book},
$\tau_{d} \propto (N - 1)^{3}$).
Thus we find that the longest relaxation time and the zero shear
is also reasonably
reproduced by our simulation model.

Figure \ref{zero_shear_viscosity} shows the zero shear viscosity
$\eta_{0}$.
The zero shear viscosity is calculated from $G(t)$ as
\begin{equation}
 \eta_{0} = \int_{0}^{\infty} dt \, G(t)
\end{equation}
We find that the viscosity is proportional to $N - 1$ or $(N - 1)^{3.40}$
for small or large $N$, respectively.
The cross over bead number is roughly estimated to be $N_{c} \approx
14.2 \approx 1.4 N_{e}$.
This value is slightly smaller than the experimental value $N_{c} / N_{0} = 1.6 \sim 3.5$
\cite{physical-properties-of-polymers-handbook-chapter25}, but the
discrepancy is not so large.
Although we do not show simulation results for other parameter sets (for
example, for different $N_{0}$, $N_{s}$, or $\zeta_{s}$), as shown in Ref
\citen{Likhtman-2005}, the effects of these parameters are not large and
we have qualitatively similar results. Thus in this work we limit
ourselves only to the standard parameter set used in Ref \citen{Likhtman-2005}.

Next we calculate the viscosity growth $\eta(t,\dot{\gamma})$ and the
steady state viscosity $\eta(\dot{\gamma})$. At $t = 0$, the system is
in equilibrium. For $t > 0$ we apply the constant shear flow as
\begin{equation}
 \kappa_{\alpha \beta}
  = \begin{cases}
     \dot{\gamma} & (\alpha = x, \beta = y) \\
     0 & (\text{otherwise})
    \end{cases}
\end{equation}
and calculate $\eta(t,\dot{\gamma})$ and $\eta(\dot{\gamma})$ as follows.
\begin{align}
 & \eta(t,\dot{\gamma}) = \frac{\langle \sigma_{xy}(t)
 \rangle}{\dot{\gamma}} \\
 & \eta(\dot{\gamma}) = \lim_{t \to \infty} \eta(t,\dot{\gamma})
\end{align}
We also calculate the dynamic viscosity, which is defined as follows, to
check whether the Cox-Merz rule \cite{Graessley-book} holds or not.
\begin{align}
 & \label{cox_merz_rule}
 \eta(\dot{\gamma}) \approx \left. \eta^{*}(\omega) \right|_{\omega
 = \dot{\gamma}} \\
 & \label{dynamic_viscosity_definition}
 \eta^{*}(\omega) \equiv \sqrt{G'^{2}(\omega) + G''^{2}(\omega)}
\end{align}

Figure \ref{viscosity_growth} shows the viscosity growth curves for $N =
40$ with various shear rates ($\dot{\gamma} \tau_{0} = 0.0025,$ $0.005,$
$0.01,$ $0.025,$ $0.05,$ and $ 0.1$). The number of sample chains is $M =
1024$ for all shear rates. 
As shown in Figure \ref{viscosity_growth},
that viscosity
growth curves can be qualitatively well reproduced by the single chain
slip-spring model. The stress overshoot behavior and the shear thinning
behavior are reasonably reproduced.

Figure \ref{dynamic_and_steady_viscosities} shows the dynamic and steady
state viscosities calculated for $N = 10,$ $20,$ $40,$ and $80$.
For relatively small $\dot{\gamma}$ and $\omega$, we observe that the
Cox-Merz rule \eqref{cox_merz_rule} holds reasonably. However,
we find that in high shear rate regions, the steady state
viscosities deviate from the complex viscosity and the Cox-Merz rule does
not hold. In such regions, the viscosities are nearly independent of
shear rate, thus they are the second Newtonian viscosities.
These second Newtonian viscosities can be understood as follows. In high
shear rate regions, slip-springs as well as chains are strongly
stretched and slip-springs are easily destructed. Thus there are only a
few slip-springs on a chain. For example, for $N = 80$ ($\langle Z
\rangle_{\text{eq}} = 20$), the steady state average number of
slip-springs are $\langle Z \rangle = 8.7, 4.2,$ and $1.6$ for
$\dot{\gamma} = 0.01, 0.1$, and $1$, respectively.
Then, a chain behaves essentially like an ideal,
Rouse chain. The viscosity of a Rouse chain is independent of shear rate
since the Rouse model is linear. Following this picture, we expect that
the second Newtonian viscosity depends on $N$ as
\begin{equation}
 \label{second_newtonian_viscosity_rouse}
 \frac{\eta(\dot{\gamma})}{\rho_{0} k_{B} T \tau_{0}} \propto N \qquad
\end{equation}
In Figure \ref{dynamic_and_steady_viscosities}, we can observe that
eq \eqref{second_newtonian_viscosity_rouse}
approximately holds at the high shear rate region in
Figure \ref{dynamic_and_steady_viscosities}.
Therefore we conclude that the second Newtonian like behavior in high
shear rate region is an artifact of our model, rather than a physical
property of an entangled chain. This means
that our model should not be applied to very high shear rate regions
where the Cox-Merz rule is violated.
Results for the steady state
viscosity may be utilized to validate simulation results under more complex
situations, such as flows around obstacles.

{ In the shear thinning region in Figure
\ref{dynamic_and_steady_viscosities}, the steady
state viscosity obeys the power law
$\eta(\dot{\gamma}) \propto \dot{\gamma}^{- \alpha}$.
The exponent $\alpha$ is smaller than $1$ and thus
the steady state shear stress $\sigma_{xy}(\dot{\gamma}) \propto
\dot{\gamma}^{1 - \alpha}$ is a
monotonically increasing function of the shear rate $\dot{\gamma}$.
In the simple Doi-Edwards type tube model\cite{Doi-Edwards-book}, $\sigma_{xy}(\dot{\gamma})$
is not a monotonically increasing function of $\dot{\gamma}$, and such a non-monotonic relation can cause
mechanical instability (shear-banding instability)
\cite{Cates-McLeish-Marrucci-1993}. It is known that the CCR mechanism
can successfully remove this instability
\cite{Mead-Larson-Doi-1998,Graham-Likhtman-McLeish-Milner-2003}.
As we mentioned, the CR or CCR mechanisms are not explicitly taken into
our model. Even in absence of the CCR mechanism, our model is free from
the mechanical instability (at least in the investigated parameter range).}

{ Finally, we shortly investigate the contribution of
the virtual stress tensor $\hat{\bm{\sigma}}^{(v)}$.
In Section \ref{linear_response_theory} we assumed that the stress
tensor of the system is given by $\hat{\bm{\sigma}}$ defined by eq
\eqref{stress_tensor_operator_single_chain}. One may prefer to employ
$\hat{\bm{\sigma}} + \hat{\bm{\sigma}}^{(v)}$ as the stress tensor of
the system, which is conjugate to the velocity gradient tensor.
Here we consider the shear relaxation modulus as an example. The
relative contribution of the virtual stress to the shear relaxation
modulus can be defined as
\begin{equation}
 \label{relative_contribution_virtual_stress_to_shear_relaxation_modulus}
  \Delta \tilde{G}^{(v)}(t)
  \equiv \frac{\langle
  \hat{\sigma}^{(v)}_{xy}(t)
  \hat{\sigma}_{xy}
  \rangle_{\text{eq}} + \langle
  \hat{\sigma}^{(v)}_{xy}(t)
  \hat{\sigma}^{(v)}_{xy} 
  \rangle_{\text{eq}}}
  {\langle
  \hat{\sigma}_{xy}(t)
  \hat{\sigma}_{xy}
  \rangle_{\text{eq}} + \langle
  \hat{\sigma}_{xy}(t)
   \hat{\sigma}^{(v)}_{xy}
  \rangle_{\text{eq}}}
\end{equation}
If $\Delta \tilde{G}^{(v)}(t)$ is negligibly small, or if it is independent
of $t$, the stress-optical rule approximately
holds even if we employ $\hat{\bm{\sigma}} + \hat{\bm{\sigma}}^{(v)}$ as
the stress tensor of the system.
Figure \ref{relative_error_shear_relaxation_modulus} shows the relative
contribution of the virtual stress for $N = 40$.
As shown in Figure \ref{relative_error_shear_relaxation_modulus},
$\Delta \tilde{G}^{(v)}(t)$ is roughly about $25\%$ and this value is not
negligibly small. However, it does not strongly depend on time $t$. This
means that if we employ $\hat{\bm{\sigma}} + \hat{\bm{\sigma}}^{(v)}$ as
the stress tensor of the system, the stress-optical rule is
approximately valid and rheological properties would be
qualitatively not changed (quantitatively they would be changed by about
$25\%$). The results shown in this subsection are therefore
qualitatively not sensitive to the definition of the stress tensor.}

Judging from obtained rheological data, we can conclude that our model
reasonably reproduces rheological properties and
can be used to study rheological properties or flow behaviors of
entangled polymers. Although further improvement of the model will be
possible, it is beyond the scope of the current work and left for a future
work.


\subsection{Comparison of Rheological Properties with Original Slip-Spring Model}

In this subsection, we briefly compare rheological data
calculated by our single chain slip-spring model and the Likthman's
original slip-spring model. We compare several rheology data shown in
Ref \citen{Likhtman-2005} with the results of our simulations.

To avoid numerical errors due to the fitting or the numerical integration, we
compare the shear relaxation modulus instead of storage and loss moduli.
Figure \ref{compare_shear_relaxation_modulus}(a) shows the shear relaxation
moduli calculated by our model and the original slip-spring model, for
$N = 8, 16, 32, 64$ and $128$. Other parameters ($N_{0}, N_{s}$ and
$\zeta_{s}$) are the same.
The data of the orignal model is taken from Fig. 4(a) of
Ref \citen{Likhtman-2005}.
We can observe that the forms of
$G(t)$ calculated by our model and the original model are quite similar for long time
region ($t \gtrsim 10 \tau_{0}$), while the longest relaxation times are
quantitatively different.
Our model gives longer relaxation time for all cases. We consider
this is mainly due to the lack of the CR effect in our model.
(There may be other reasons for this discrepancy,
such as the effect of the short time scale dynamics.
However we consider their contributions are not large
compared with the CR effect.)

Nonetheless, the relaxation behavior of our model is qualitatively
similar to the original model with the CR effect. To see it clearly,
we show the shear relaxation moduli shifted
vertically and horizontally in Figure
\ref{compare_shear_relaxation_modulus}(b). The data by our model (solid
curves in Figure \ref{compare_shear_relaxation_modulus}(a)) are shifted
(the data by the original model are not shifted).
For relatively long time region ($t \gtrsim \tau_{0}$), our data can be
collapsed to the data obtained by the original model well.
The horizontal and vertical shift factors, $\tilde{a}$ and $\tilde{b}$,
are of the order of unity ($\tilde{a} \approx 1.9$ and $\tilde{b} \approx 0.8$
for $N = 128$), and
they slightly depend on $N$. This will be due to the difference between
the dependeces of relaxation mechanisms to $N$.
%

Figure \ref{compare_zero_shaer_viscosities} shows zero shear viscosities
calculated by our model and the original model (with or without the CR
effect).
The data of the orignal model is
taken from Fig. 5(a) of Ref \citen{Likhtman-2005}.
We can observe that for large $N$, $\eta_{0}$ by our model is close to
the one by
the original model without the CR. On the other hand, for small $N$, our
model gives smaller $\eta_{0}$ compared with the original model.
We consider this is due to the difference of the dynamics of
slip-springs.
In our model, slip-springs can be attatched only on beads while in the
original model, slip-springs can be attatched in between beads. Besides,
our model allows slip-springs to pass through each other.
The slip-spring dynamics of our model seems to give faster relaxation
for small $N$, compared with the original model.

From the above comparisons for $G(t)$ and $\eta_{0}$, we conclude that our model
can reproduce rheological properties of the original model qualitatively
well, especially for well entangled polymers.
The relaxation time or the zero shear viscosity are close to the
data by the original model without the CR effect. This is natural since
our model does not incorporate the CR effect.
Nonetheless, the relaxation behavior 
of our model is almost the same as one of the original model in the long time region ($t \gtrsim 10
\tau_{0}$).

\subsection{Acceleration by GPU}

In the previous subsection, we have shown that the single chain
slip-spring model can reproduce rheological properties qualitatively.
In this subsection we show the results of the acceleration by a GPU.
To study the acceleration, we compare the calculation times on a CPU and
on a GPU, by using the same parameter set. As an example, in this work
we use Intel Core 2 Duo E8500 ($3.16\text{GHz}$, dual-core) for
simulations on CPU, and NVIDIA Tesla C1060 ($1.3\text{GHz}$, $240$ CUDA cores) for
simulations on GPU. Simulation programs are written in C and CUDA for
CPU and GPU, respectively. They are compiled by using gcc (version
4.3.0) and nvcc (version 2.1) and executed on Linux (kernel 2.6.9, x86\_64).
The program for CPU is written in ANSI C and no CPU-specific
extensions (such as the SSE or SSE2 instruction sets
\cite{ia32-software-developers-manual-vol1}) are utilized.
The numbers of threads used for the calculations are $1$ (CPU) and $128
\times 128$ (GPU).
For comparison, we performed simulations with the following
parameters; the numbers of beads and chains $N \times M = 16 \times 4096,32 \times 2048,64 \times 1024,$
or $128 \times 512$ (the total number of beads is kept to be constant),
the shear rate $\dot{\gamma} = 0$ or $0.05$. All the simulations are
started from the equilibrium initial state (at time $t = 0$).
The time
step size is $\Delta t = 0.01$, and the simulations are stopped at time
$t = 100$ (the total number of time steps is $10000$).
The results are summarized in Table \ref{calculation_times}.
We can observe that the program for GPU is about $290$ times faster than the
program for CPU. The acceleration by a GPU is quite
effective to accelerate our single chain slip-spring model.

\section{Discussion}

\subsection{Model Properties}

We formulated a single chain version of the
slip-spring model for entangled polymers. Our model is designed as the
simplified model of the original one. One notable property of our model
is that it fully satisfies the detailed balance condition. This means
that the equilibrium probability distribution rigorously becomes the
Boltzmann distribution. This enables us to tune the equilibrium
statistics of the model easily. For example, we can employ the
statistics of a non-ideal chain (real chain) for our model, or we can
introduce the interaction between slip-springs. Such modifications can be
done essentially only by changing the expression for the grand potential
\eqref{grand_potential_single_chain}.

Because it is reported that the statistics
of entangled polymer chains somehow depend on models (such as primitive
path extraction methods
\cite{Everaers-Sukumaran-Grest-Svaneborg-Sivasubramanian-Kremer-2004,Kroger-2005,Tzoumanekas-Theodorou-2006}
or dynamic equations
\cite{Masubuchi-Uneyama-Watanabe-Ianniruberto-Greco-Marrucci-2010}), it
will be desirable for a model to be tunable for a specific target
statistics.
The equilibrium distribution function
\eqref{full_equilibrium_distribution_single_chain_modified} has
a rather simple structure and we can tune, for example, the statistics of
the chain or the statistics slip-springs easily.
We expect that our model can be used to investigate
rheological behaviors for various chain statistics models numerically.

The detailed balance condition also becomes important when we derive the
linear response formula
\eqref{relaxation_modulus_tensor_green_kubo_form}.
Although the response formula
\eqref{relaxation_modulus_tensor_green_kubo_form} itself is already 
proposed by Ramirez, Sukumaran
and Likhtman \cite{Ramirez-Sukumaran-Likhtman-2007}, they did not give
the derivation based on the master equation.
We gave the rigorous derivation based on the master equation, which
corresponds to the Langevin and jump dynamics actually used in the simulations. Our result justifies the use of 
the response formula \eqref{relaxation_modulus_tensor_green_kubo_form}
to calculate the relaxation modulus tensor.

Although dynamics of slip-springs is modeled as a simple jump dynamics
in our model, it can reproduce linear and nonlinear rheological
behaviors qualitatively. The linear rheological properties are similar
to the original slip-spring model. This implies that our model captures
essential nature of the original slip-spring model.
We also performed simulations for nonlinear rheological properties, and
reproduced the viscosity growth and the Cox-Merz rule.
Thus we consider that our single chain slip-spring model can be used as long as
we want to calculate simple rheological properties.

In our model, the rheological properties can be reproduced well while
there is no CR effect. As we already pointed, even without the CR effect,
there is a CR like relaxation mechanism in our model (as shown in Appendix
\ref{constraint_release_type_effect}). This relaxation mechanism is
caused by the model property that slip-springs can pass through
(or exchange) each other. Our result implies that in some situations,
this ``constraint exchange'' mechanism can be employed instead of the CR
mechanism. From the numerical point of view, if we allow slip-spring to
pass through each other, the implementation becomes much easier (this is
because the time evolution of each slip-springs can be evaluated in
parallel). It seems to not be difficult to make slip-springs
exchangeable in other slip-link type models. The exchangeable slip-links
will improve numerical accuracy efficiently.

\subsection{Acceleration by GPU}

We observed that the acceleration by a GPU can improve the simulation
speed drastically. The program for GPU is about $290$ times faster than the
program for CPU, which seems to be quite efficient and promising.
However, it is fair to mention about the possible acceleration by some
CPU-specific extensional instructions.
Several CPU-specific special extensional instructions can
improve the performance largely. For example, the SSE instruction
handles 4 single precision floating-point operations in
parallel \cite{ia32-software-developers-manual-vol1}. The use of the SSE and/or SSE2 will improve the
performance roughly about 10 times (for single precision floating-point number
operations). Besides, currently the program for CPU is
not parallelized. Because we can achieve very high parallelism for
a single chain type model, the performance can be further improved by the
factor $2$ (for a dual-core CPU).
This means, even if we tune the program for
CPU extremely and parallelize it,
the GPU program is still about $15$ times faster than the
CPU program. (We also note that we can use multiple GPUs in parallel,
and it can also improve the performance.)

Although it is possible to improve the program for CPU,
typically programs with such special instructions become
quite complicated. The portability of the program is also decreased if
we use CPU-specific instructions explicitly.
As a result, the programming cost for CPU becomes much larger than
one for GPU.
Judging from the acceleration effects and the programming costs,
we can conclude that the single chain slip-spring model simulations on
a GPU are very efficient and promising for practical purposes.

Since simulations on a GPU enable us to calculate rheological properties
(such as stress tensor) very efficiently, in principle we can perform
{ CONNFFESSIT or particle type multiscale simulations
\cite{Laso-Ottinger-1993,Halin-Lielens-Keunings-Legat-1998,Murashima-Taniguchi-2010}} with reasonable calculation
costs, by combining our single chain slip-spring model
and macroscale fluid models. {When we perform
macroscopic fluid simulations in which many mesoscopic rheological
simulations are embedded (typically several thousand mesoscopic
simulators are embedded in a single fluid element), the mesoscale
simulations are the most time-consuming part. If we perform
such multiscale simulations only on CPUs, the required calculation time is still
considerable even if we parallelize the
mesoscopic simulator. Our simulation model and use of a GPU can decrease
the mesoscopic calculation cost drastically.
We expect
that the total simulation time of multiscale simulations can
be also reduced drastically. Besides, the mesoscopic rheological
simulations can be further accelearated by using multiple GPUs
(because our simulations are already highly parallelized).}
Cooperating our model with macroscale fluid models will be future
works.

To study rheological properties of complicated systems, such as the rheology
of branched polymers or polymer blends, we will need to refine our
model.
We will be also required to take into account of
the CR or the CCR, for precise
calculations under fast shear rates.
For more complex architectures, such as star polymers or comb polymers,
the generalization of the model will be required.
We consider that generalization itself is not so difficult, but it may
be difficult to implement it for a GPU because there are several
limitations for a GPU. To perform simulations efficiently on a GPU, we
will need to design a generalized model so that it is suitable for
calculations on a GPU.

\section{Conclusion}

In this work we proposed a single chain slip-spring model, which is
based on the Likhtman's slip-spring model \cite{Likhtman-2005}. The model is designed to be
suitable for simulations on a GPU. Besides, the model is expressed by
using the free energy and satisfies the detailed balance condition, which
ensures that the system relaxes to the thermal equilibrium state.
We calculated several static properties (equilibrium distirbution
functions) analytically. We also calculated
the linear response of the system to strain deformation, and obtained the
Green-Kubo type formula for the relaxation modulus which is in agreement
with the one previously proposed by Ramirez,
Sukumaran and Likhtman \cite{Ramirez-Sukumaran-Likhtman-2007}.

We calculated several rheological properties such as the linear
viscoelasticity or the viscosity growth, and shown that our model can
reproduce them reasonably.
To accelerate the simulations, we performed
simulations on a GPU as well as simulations on a CPU. By comparing the
simulation times, we found that the use of a GPU can accelerate a
simulation approximately $290$ times faster. The modification of our
model or the application to actual multiscale simulations will be future
works.

\section*{Acknowledgment}

This work is supported by JST-CREST. The author thanks Mr. Ryuji
Sakamaki for informing the author about the fixed-point real number technique
and Ref \citen{Narumi-Sakamaki-Kameoka-Yasuoka-2008}.
He also thanks to Prof. Yuichi Masubuchi for various helpful comments.

\appendix

\section{Constraint Release Type Relaxation Mechanism}
\label{constraint_release_type_effect}

In many models for entangled polymers, the constraint release (CR)
effect is considered to be an important effect (especially for branched polymers).
Several methods have been developed to take the CR events into
account. Rubinstein and Colby \cite{Rubinstein-Colby-1988} modelled the CR events as hopping motions
of tube segments in a self-consistent way.
Based on this idea, Schieber and coworkers
\cite{Hua-Schieber-1998,Schieber-Neergaard-Gupta-2003,Nair-Schieber-2006}
introduced (relatively) slow diffusion type Brownian motion of
slip-links as the CR events. Masubuchi and coworkers \cite{Masubuchi-Takimoto-Koyama-Ianniruberto-Greco-Marrucci-2001}
directly modelled the CR
events as reconstruction of slip-links between two chains.
Doi and Takimoto \cite{Doi-Takimoto-2003}, and Likhtman
\cite{Likhtman-2005} modelled the CR events as
reconstruction events in a similar way (in their models, slip-links
(slip-springs) are virtually paired).
In our single chain slip-spring model, the CR process is not explicitly
considered.
However, as we discuss in this section, the CR type relaxation process
exists (implicitly) in our model.

One peculiar property of our model is that slip-springs can pass through
each other, unlike the original slip-spring model. Most of
lip-link based models do not allow slip-links to pass through each
other. Similarly, it is usually not allowed in most of tube models to exchange the
neighboring entanglement points. Then, we can expect that
the passing-through events of slip-springs will result in a sort of
relaxation process.

We consider a passing-through event of two neighboring
slip-springs.
Here we label the slip-spring indices $j$ in the following
order (ascending in $S_{j}$) to compare our model with conventional models.
\begin{equation}
 \label{bead_indices_order_ascending}
  S_{1} \le S_{2} \le S_{3} \le \dots \le S_{Z}
\end{equation}
This condition is (implicitly) assumed in many
slip-link based models.
If the $j$-th and $(j + 1)$-th slip-springs are exchanged at time $t$,
then we should exchange the slip-spring indices and anchoring points to
satisfy the condition \eqref{bead_indices_order_ascending}.
\begin{align}
 & S_{j}(t + 0) = S_{j + 1}(t), \qquad S_{j + 1}(t + 0) = S_{j}(t) \\
 & \bm{A}_{j}(t + 0) = \bm{A}_{j + 1}(t), \qquad \bm{A}_{j + 1}(t + 0) = \bm{A}_{j}(t)
\end{align}
This dynamics of two monomer indices, $S_{j}$ and $S_{j + 1}$, can be
interpreted as the collision and reflection like dynamics.
The dynamics of two anchoring points, $\bm{A}_{j}$ and $\bm{A}_{j
+ 1}$, can be interpreted as sudden jumps in space. We consider that
such jump events are similar to the CR picture considered by Rubinstein and Colby \cite{Rubinstein-Colby-1988}.
Thus, we can interpret the exchange events as the
CR like motions of anchoring points. Then the exchange events
effectively give the CR like stress relaxation process.

However, we should notice that the exchange events do not exactly
correspond to the conventional CR events. For example, our CR like
events are non-Markovian while the CR events are usually modelled as
Markovian. (This is because after one exchange event, the same
slip-spring pair can pass through each other again. Such a process results
in the memory effect.)
Thus the effect of the exchange events to the stress relaxation is
expected not to be strong at long time scale.
To fully take account of the conventional CR
events, we will need to model the CR process in our model as another
jump process.

\section{Detailed Calculations in Linear Response Theory}
\label{detail_calculations_in_linear_response_theory}

In this appendix, we show detailed calculations in the
derivation of the linear response of the stress to the velocity gradient tensor.
Although the calculations themselves are rather straightforward, the
final result is not so intuitive. We show detailed calculations to avoid
confusions. We note that, following the same procedure, one can derive
other linear response functions such as the dielectric response
function. Mainly we follow the standard derivation of the linear
response theory for the Fokker-Planck equation \cite{Risken-book}.

We may start from eq
\eqref{time_evolution_probability_distribution_single_chain_modified},
the time evolution equation for the probability distribution function.
If we decompose the probability distribution function into the
equilibrium and perturbation parts,
\begin{equation}
 P(\lbrace \bm{r}_{i} \rbrace,\lbrace \bm{a}_{j} \rbrace,\lbrace s_{j}
  \rbrace,z;t)
  =  P_{\text{eq}}(\lbrace \bm{r}_{i} \rbrace,\lbrace \bm{a}_{j} \rbrace,\lbrace s_{j}
  \rbrace,z) + P_{1}(\lbrace \bm{r}_{i} \rbrace,\lbrace \bm{a}_{j} \rbrace,\lbrace s_{j}
  \rbrace,z;t)
\end{equation}
the time evolution equation
\eqref{time_evolution_probability_distribution_single_chain_modified} can be approximately expressed as follows,
by taking only the linear terms in the perturbation expansion.
\begin{equation}
 \label{time_evolution_probability_distribution_single_chain_modified_linear}
 \frac{\partial P_{1}}{\partial t}
 \approx {\mathcal{L}}_{0} P_{1} + {\mathcal{L}}_{1}(t) P_{\text{eq}}
\end{equation}
where we used that the equilibrium part of the time evolution operator, ${\mathcal{L}}_{0}$, and 
the equilibrium distribution function $P_{\text{eq}}$ satisfy the
following equation.
\begin{equation}
 \frac{\partial P_{\text{eq}}}{\partial t} = {\mathcal{L}}_{0} P_{\text{eq}} = 0
\end{equation}

By integrating eq
\eqref{time_evolution_probability_distribution_single_chain_modified_linear}
we have
\begin{equation}
 \label{time_evolution_probability_distribution_single_chain_modified_linear_integrated}
 P_{1}(\lbrace \bm{r}_{i} \rbrace,\lbrace \bm{a}_{j} \rbrace,\lbrace s_{j}
  \rbrace,z;t)
  = \int_{-\infty}^{t} dt' \, e^{(t - t') {\mathcal{L}}_{0}}
  {\mathcal{L}}_{1}(t') P_{\text{eq}}
\end{equation}
The average time-dependent stress tensor is then calculated to be
\begin{equation}
 \label{average_stress_tensor_single_chain}
 \begin{split}
 \bm{\sigma}(t) & = \sum_{z,\lbrace s_{i} \rbrace}
  \int d\lbrace \bm{r}_{i} \rbrace d\lbrace \bm{a}_{j} \rbrace \,
  \hat{\bm{\sigma}}  P(\lbrace \bm{r}_{i} \rbrace,\lbrace \bm{a}_{j} \rbrace,\lbrace s_{j}
  \rbrace,z;t) \\
  & = \sum_{z,\lbrace s_{i} \rbrace}
  \int d\lbrace \bm{r}_{i} \rbrace d\lbrace \bm{a}_{j} \rbrace \,
  \hat{\bm{\sigma}} \left[ P_{\text{eq}}
  + \int_{-\infty}^{t} dt' \, e^{(t - t') {\mathcal{L}}_{0}}
  {\mathcal{L}}_{1}(t') P_{\text{eq}}
  \right] \\
  & = \bm{\sigma}_{\text{eq}}
  + \int_{-\infty}^{t} dt' \, \sum_{z,\lbrace s_{i} \rbrace}
  \int d\lbrace \bm{r}_{i} \rbrace d\lbrace \bm{a}_{j} \rbrace \, \hat{\bm{\sigma}} e^{(t - t') {\mathcal{L}}_{0}}
  {\mathcal{L}}_{1}(t') P_{\text{eq}}
 \end{split}
\end{equation}
where $\bm{\sigma}_{\text{eq}} \equiv \langle \hat{\bm{\sigma}} \rangle_{\text{eq}}$.
Eq \eqref{average_stress_tensor_single_chain} can be modified further by
utilizing the following equation.
\begin{equation}
 \begin{split}
 {\mathcal{L}}_{1}(t) P_{\text{eq}}
  & = \frac{1}{k_{B} T} \bigg[ \sum_{i = 1}^{N} \left[ 
  \frac{\partial \mathcal{J}}{\partial \bm{r}_{i}} \bm{r}_{i}
  - k_{B} T \bm{1} \right]
  + \sum_{j = 1}^{z} \left[
  \frac{\partial \mathcal{J}}{\partial \bm{a}_{j}} \bm{a}_{j} - k_{B} T
  \bm{1} \right] \bigg] : \bm{\kappa}(t) P_{\text{eq}} \\
  & = \frac{1}{k_{B} T} \big[ \hat{\bm{\sigma}} +
  \hat{\bm{\sigma}}^{(v)} \big] : \bm{\kappa}(t) P_{\text{eq}}
 \end{split}
\end{equation}
where we defined the virtual stress tensor operator $\hat{\bm{\sigma}}^{(v)}$as
\begin{equation}
 \hat{\bm{\sigma}}^{(v)} \equiv \sum_{j = 1}^{z} \frac{3 k_{B} T}{N_{s} b^{2}} (\bm{r}_{s_{j}} -
  \bm{a}_{j})
  (\bm{r}_{s_{j}} - \bm{a}_{j})
 - z k_{B} T \bm{1}
\end{equation}
Finally we have the following expression for the time-dependent stress
tensor, and thus we have eq \eqref{single_chain_time_depending_stress_expression_final}.
\begin{equation}
 \label{single_chain_time_depending_stress_expression_virtual_stress}
 \begin{split}
 \bm{\sigma}(t)
  & = \bm{\sigma}_{\text{eq}}
  + \frac{1}{k_{B} T} \int_{-\infty}^{t} dt' \, \sum_{z,\lbrace s_{i} \rbrace}
  \int d\lbrace \bm{r}_{i} \rbrace d\lbrace \bm{a}_{j} \rbrace \, \hat{\bm{\sigma}} e^{(t - t') {\mathcal{L}}_{0}}
  \left[ \big[ \hat{\bm{\sigma}} +
  \hat{\bm{\sigma}}^{(v)} \big] : \bm{\kappa}(t') P_{\text{eq}} \right] \\
  & = \bm{\sigma}_{\text{eq}}
  + \frac{1}{k_{B} T} \int_{-\infty}^{t} dt' \, \sum_{z,\lbrace s_{i} \rbrace}
  \int d\lbrace \bm{r}_{i} \rbrace d\lbrace \bm{a}_{j} \rbrace \,
  \left[ e^{(t - t') {\mathcal{L}}_{0}^{\dagger}} \hat{\bm{\sigma}} \right]
  \big[ \hat{\bm{\sigma}} +
  \hat{\bm{\sigma}}^{(v)} \big] P_{\text{eq}} : \bm{\kappa}(t') \\
  & = \bm{\sigma}_{\text{eq}}
  + \frac{1}{k_{B} T} \int_{-\infty}^{t} dt' \, \big\langle
  \hat{\bm{\sigma}}(t - t')
  \big[ \hat{\bm{\sigma}} +
  \hat{\bm{\sigma}}^{(v)} \big] \big\rangle_{\text{eq}} : \bm{\kappa}(t')
 \end{split}
\end{equation}
Here ${\mathcal{L}}_{0}^{\dagger}$ is the adjoint operator of
${\mathcal{L}}_{0}$, which is defined
via the following relation.
\begin{equation}
 \sum_{z,\lbrace s_{i} \rbrace}
  \int d\lbrace \bm{r}_{i} \rbrace d\lbrace \bm{a}_{j} \rbrace \,
  \hat{B} {\mathcal{L}}_{0} P_{\text{eq}}
  =  \sum_{z,\lbrace s_{i} \rbrace}
  \int d\lbrace \bm{r}_{i} \rbrace d\lbrace \bm{a}_{j} \rbrace \,
  ({\mathcal{L}}_{0}^{\dagger} \hat{B}) P_{\text{eq}}
\end{equation}
where $\hat{B}$ is an arbitrary operator. We also defined the
time-shifted operator of $\hat{B}$ as follows, from the fact that
$\mathcal{L}_{0}^{\dagger}$ works as the equilibrium time evolution operator.
\begin{equation}
 \label{time_shifted_operator_definition}
 \hat{B}(t) \equiv e^{t {\mathcal{L}}_{0}^{\dagger}} \hat{B}
\end{equation}
where $\hat{B}$ is again an arbitrary operator.

We note that the
detailed balance condition is essential in the preceding derivation of the linear
response formula.
If the model does not satisfy the detailed balance, generally
the simple Green-Kubo type formulae such as eq
\eqref{relaxation_modulus_tensor_green_kubo_form} do not hold.
Although we can calculate
the linear responses even if the detailed balance condition is not
satisfied, generally the resulting expressions
do not reduce to the Green-Kubo form.

\bibliographystyle{srj}
\bibliography{polymer,langevin,accelerator,tmp}

\clearpage

\section*{Figure and Table Captions}

Figure \ref{storage_and_loss_moduli}: Storage and loss moduli calculated by the single chain
slip-spring model. Circles and crosses indicate $G'(\omega)$
and $G''(\omega)$ calculated by simulations.

\

\hspace{-\parindent}%
Figure \ref{longest_relaxation_time}: The longest relaxation time calculated from the linear
 viscoelasticity data. Broken lines show slopes 2, 3 and 3.48, which
 correspond to the Rouse type relaxation, the pure reptation type
 relaxation, and the exponent obtained by the fitting, respectively.

\

\begin{figure}[htb]
 \centering
 {\includegraphics[width=0.95\linewidth,clip]{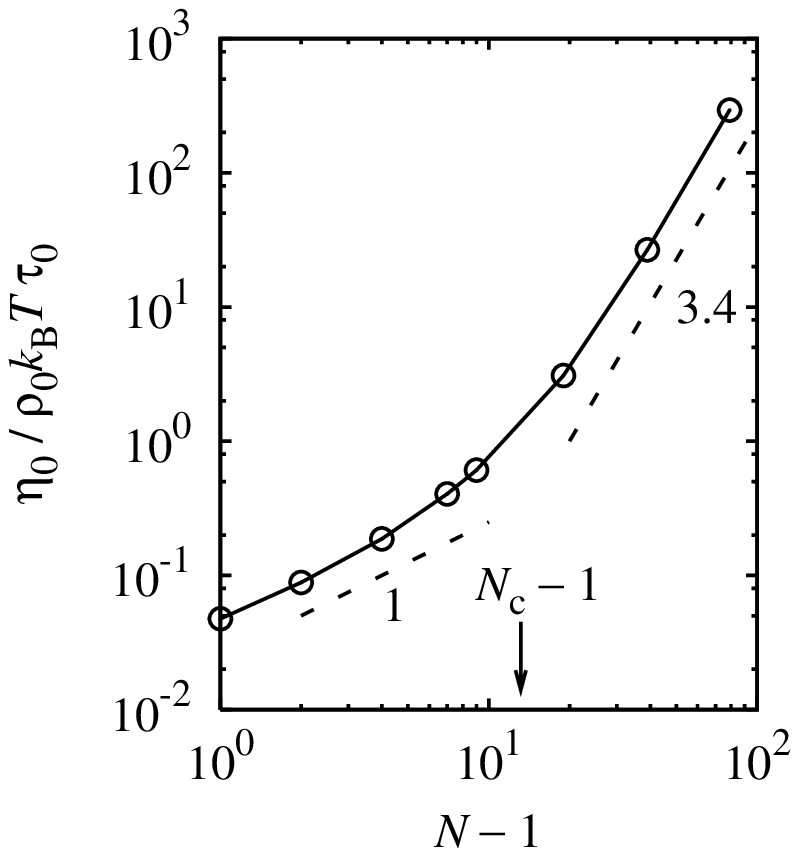}}
 \caption{\label{zero_shear_viscosity}}
\end{figure}

\begin{figure}[htb]
 \centering
 {\includegraphics[width=0.95\linewidth,clip]{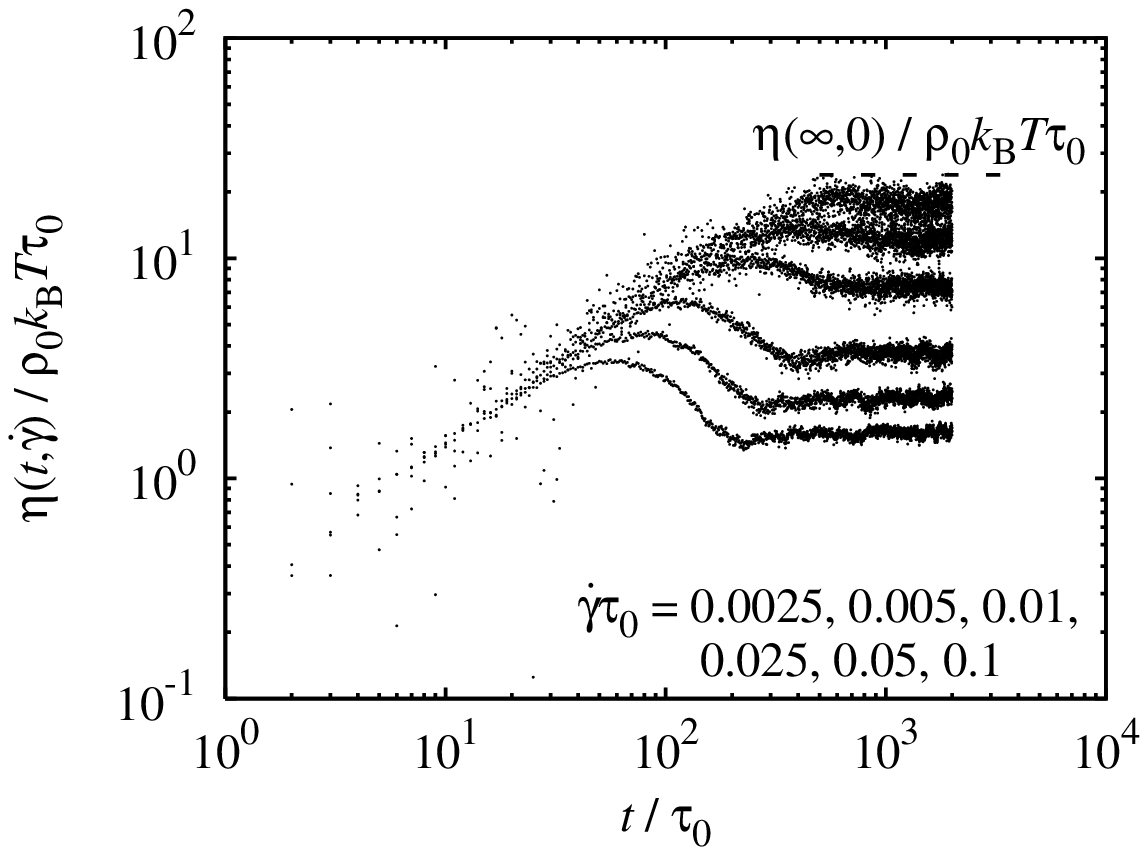}}
 \caption{\label{viscosity_growth}}
\end{figure}

\begin{figure}[htb]
 \centering
 {\includegraphics[width=0.95\linewidth,clip]{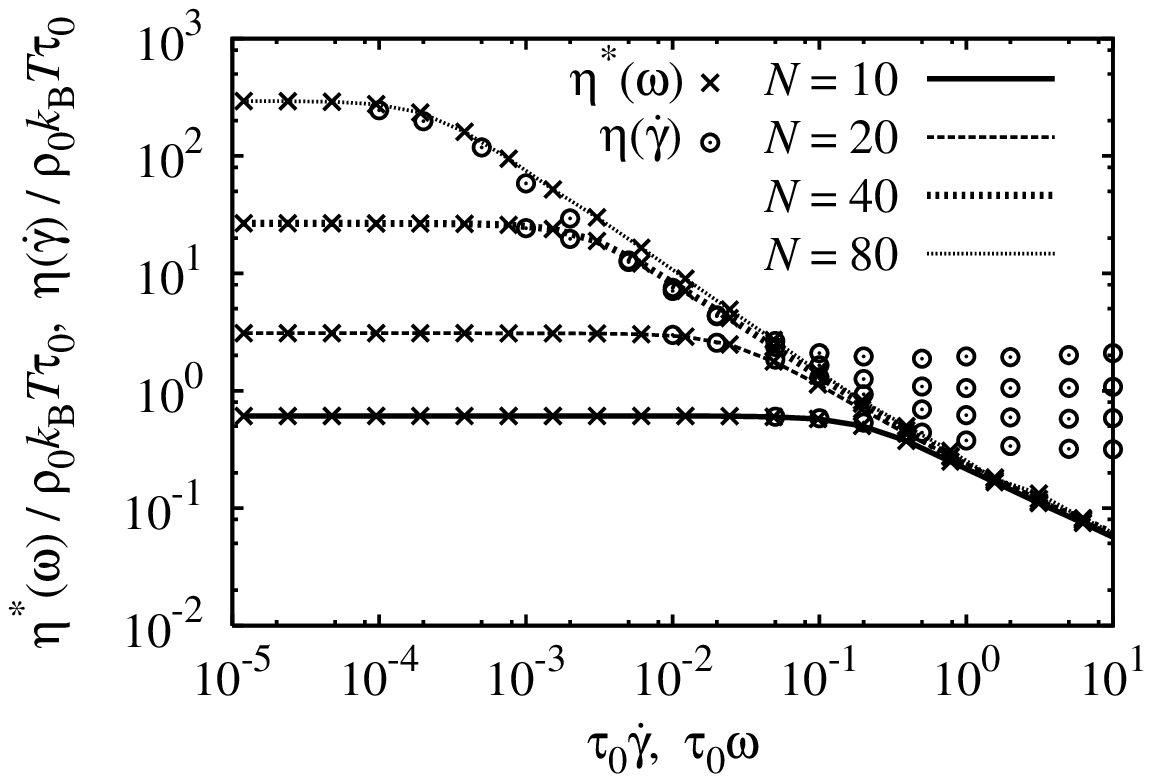}}
 \caption{\label{dynamic_and_steady_viscosities}}
\end{figure}

\begin{figure}[htb]
 \centering
 {\includegraphics[width=0.8\linewidth,clip]{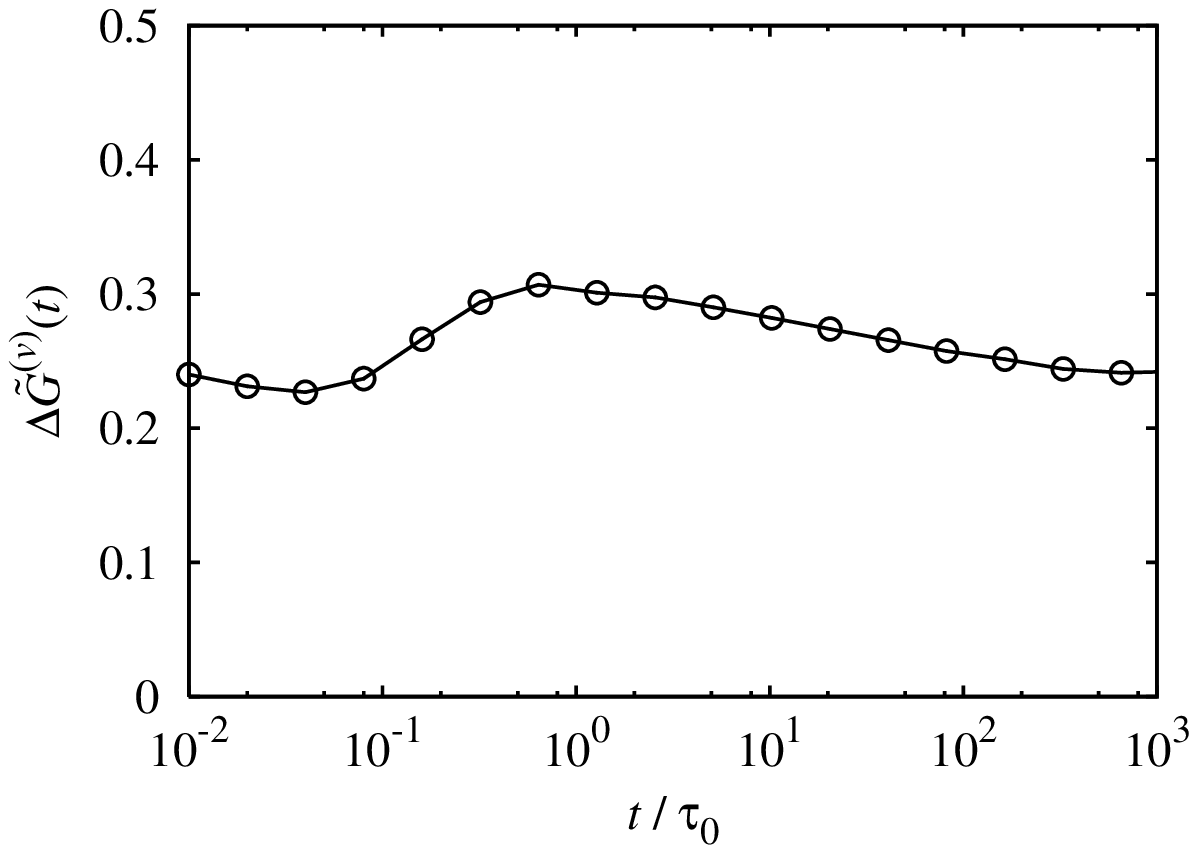}}
 \caption{\label{relative_error_shear_relaxation_modulus}}
\end{figure}

\begin{figure}[htb]
 \centering
 {\includegraphics[width=0.95\linewidth,clip]{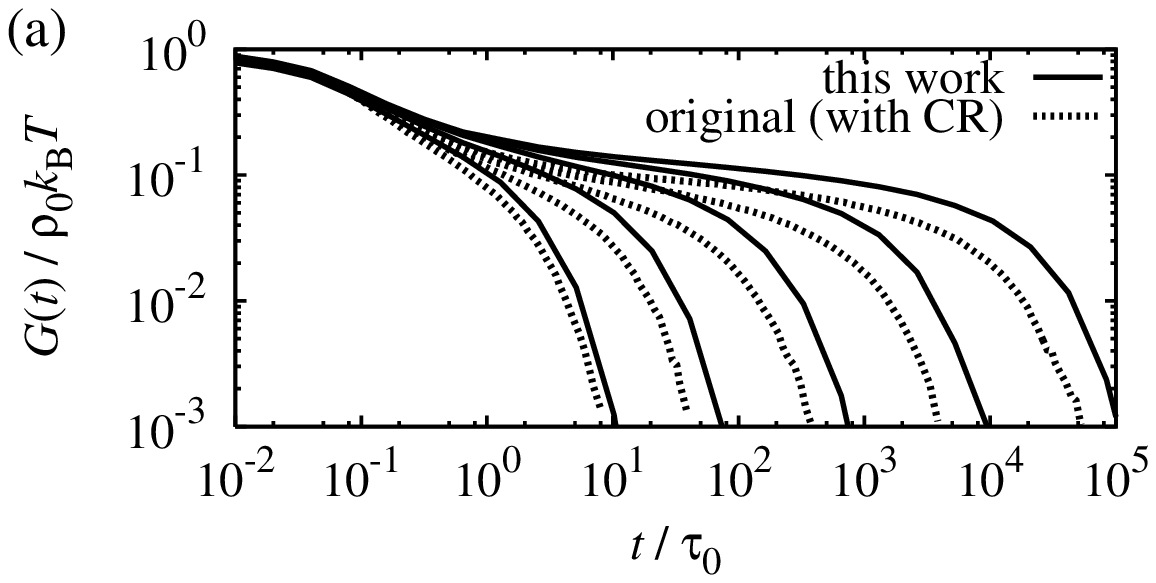}}
 {\includegraphics[width=0.85\linewidth,clip]{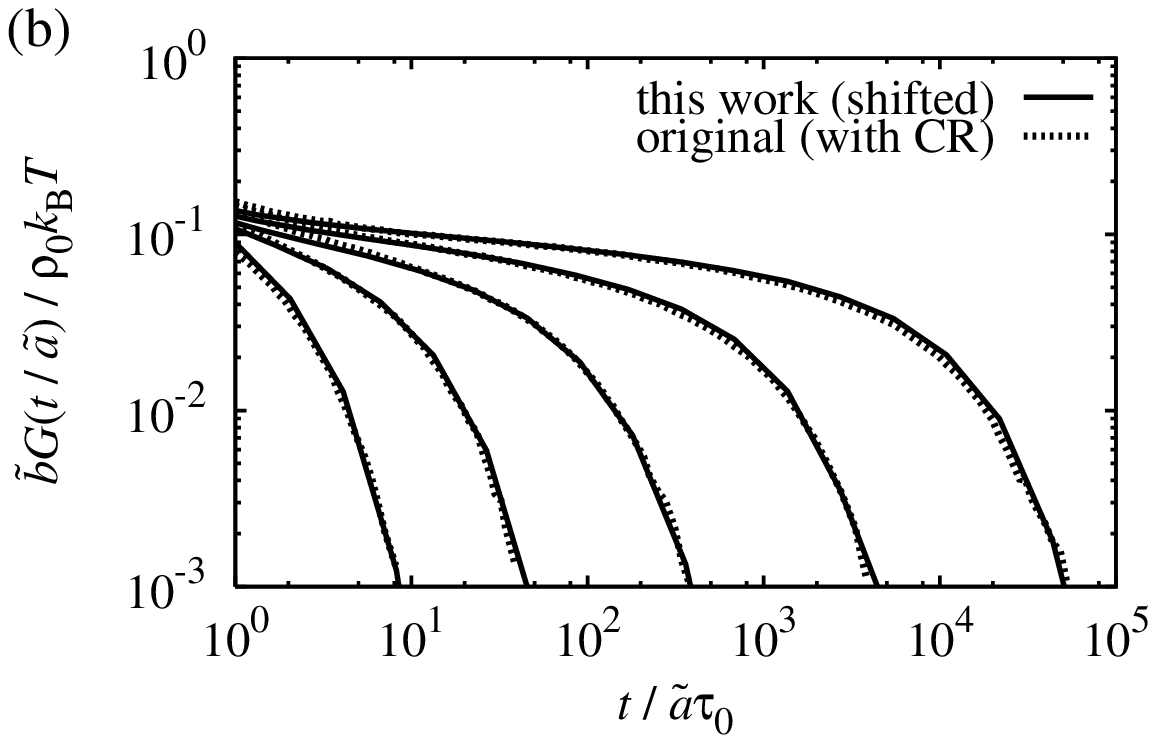}}
 \caption{\label{compare_shear_relaxation_modulus}}
\end{figure}

\begin{figure}[htb]
 \centering
 {\includegraphics[width=0.95\linewidth,clip]{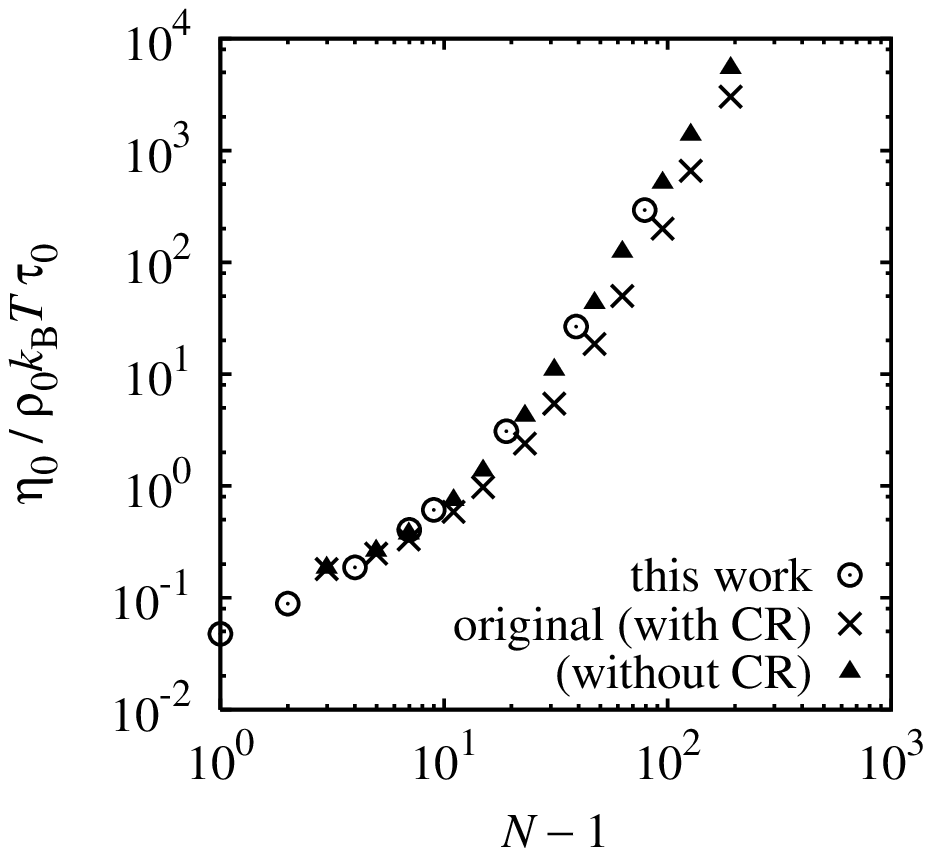}}
 \caption{\label{compare_zero_shaer_viscosities}}
\end{figure}

\hspace{-\parindent}%
Figure \ref{zero_shear_viscosity}: Zero shear viscosity calculated from the linear
 viscoelasticity data. Broken lines show slopes 1 and 3.4. The critical
 bead number is estimated to be $N_{c} = 14.2$. (The arrow indicates the
 critical bead number.)

\

\hspace{-\parindent}%
Figure \ref{viscosity_growth}: Viscosity growth curves with various shear rates for $N = 40$.
 The shear rates are $\dot{\gamma} \tau_{0} = 0.0025,$ $0.005,$ $0.01,$
$0.025,$ $0.05,$ and $0.1$.
 The broken line shows the zero shear viscosity calculated from the
 linear viscoelasticity data.

\

\hspace{-\parindent}%
Figure \ref{dynamic_and_steady_viscosities}: Dynamic and steady viscosities, $\eta^{*}(\omega)$ and $\eta(\dot{\gamma})$.
 Circles indicate $\eta^{*}(\omega)$ and the crosses (with curves)
 indicate the $\eta(\dot{\gamma})$.

\

\hspace{-\parindent}%
{ Figure \ref{relative_error_shear_relaxation_modulus}: Relative
contribution of the virtual stress to the shear relaxation modulus.}

\

\hspace{-\parindent}%
Figure \ref{compare_shear_relaxation_modulus}: (a) Comparison of shear
relaxation moduli calculated by original and our slip-spring models.
Solid curves are calculated by our model.
Dotted curves are the data of the original slip-spring model with or without
the CR effect (taken from Fig. 4(a) of Ref \citen{Likhtman-2005}).
$N = 8, 16, 32, 64$ and $128$, from left to right. (b) The same data
with (a) but data calculated by our model are shifted horizontally and
vertically.

\

\hspace{-\parindent}%
Figure \ref{compare_zero_shaer_viscosities}: Comparison of zero shear viscosities
calculated by original and our slip-spring models. Circles are
calculated by our model (and the same as the data shown in Figure
\ref{zero_shear_viscosity}).
Crosses and triangles are the data of the original slip-spring model with
and without CR (taken from Fig. 5(a) of Ref \citen{Likhtman-2005}).

\

\hspace{-\parindent}%
Table \ref{calculation_times}: Calculation times with various simulation parameters on a
 CPU and on a GPU.
 $N,M,\dot{\gamma}$ are the number of beads per chain (polymerization
 index), the total number of chains, and the shear rate.
 $t_{\text{CPU}}$ and $t_{\text{GPU}}$ are the calculation times on a
 CPU (Intel Core 2 Duo E8500, single thread) and on a GPU (NVIDIA Tesla
 C1060, $128 \times 128$ threads),
 respectively. Simulations are performed from 
 $t = 0$ (in equilibrium) to $t = 100$ with the time step
 size $\Delta t = 0.01$.
 The speed-up factor is defined as the ratio of calculation times,
 $t_{\text{CPU}} / t_{\text{GPU}}$.

\clearpage

\section*{Figures}

\clearpage

\section*{Table}

\begin{table}[htb]
 \begin{center}
 \begin{tabular}{cccccc}
  \hline
  $N \times M$ & $\dot{\gamma}$ & $t_{\text{CPU}} \text{[s]}$ &
  $t_{\text{GPU}} \text{[s]}$ & speed-up \\
  \hline  
  $16 \times 4096$ & $0$ & $2.66 \times 10^{2}$ & $ 8.97 \times 10^{-1}$ & $297$ \\
  $16 \times 4096$ & $0.05$ & $2.65 \times 10^{2}$ & $ 8.95 \times 10^{-1}$ & $296$ \\
  $32 \times 2048$ & $0$ & $2.58 \times 10^{2}$ & $9.02 \times 10^{-1}$ & $286$ \\
  $32 \times 2048$ & $0.05$ & $2.57 \times 10^{2}$ & $ 8.95 \times 10^{-1}$ & $287$ \\
  $64 \times 1024$ & $0$ & $2.52 \times 10^{2}$ & $8.70 \times 10^{-1}$ & $291$ \\
  $64 \times 1024$ & $0.05$ & $2.51 \times 10^{2}$ & $8.63 \times 10^{-1}$ & $292$ \\
  $128 \times 512$ & $0$ & $ 2.51 \times 10^{2}$ & $ 8.45 \times 10^{-1}$ & $297$ \\
  $128 \times 512$ & $0.05$ & $ 2.50 \times 10^{2}$ & $ 8.40 \times 10^{-1}$ & $298$ \\
  \hline
 \end{tabular}
 \end{center}
 \caption{\label{calculation_times}}
\end{table}

\end{document}